\documentclass{revtex4-2}
\usepackage[utf8]{inputenc}
\usepackage[english]{babel}
\usepackage{amsmath,theorem}
\usepackage{lipsum}
\usepackage{xcolor}
\usepackage{url}
\usepackage{subfig}
\usepackage{cancel}
\usepackage{graphicx}
\usepackage[colorlinks=true,citecolor=red,urlcolor=blue,linkcolor=blue]{hyperref}

\newcommand{\tmop}[1]{\ensuremath{\operatorname{#1}}}

\newcounter{nnacknowledgments}

{\theorembodyfont{\rmfamily}\newtheorem{acknowledgments*}[nnacknowledgments]{Acknowledgments}}

\begin{document}

\selectlanguage{english}

\title{Mono-Higgs and Mono-$Z$ Production in the Minimal Vector Dark Matter
Model}

\author{Gonzalo Benítez-Irarrázabal}
\email[Email: ]{gonzalo.benitezi@usm.cl}

\author{Alfonso R. Zerwekh}
\email[Email: ]{alfonso.zerwekh@usm.cl}
\affiliation{Departamento de F{\'i}sica, Universidad T{\'e}cnica Federico
Santa Mar{\'i}a, Chile\\
Centro Cient{\'i}fico-Tecnol{\'o}gico de Valpara{\'i}so, Universidad
T{\'e}cnica Federico Santa Mar{\'i}a, Chile\\
Millennium Institute for Subatomic physics at high energy frontier - SAPHIR,
Santiago, Chile}

\begin{abstract}
  The Minimal Vector Dark Matter is a viable realization of the minimal dark
  matter paradigm. It extends the Standard Model by the inclusion of a vector
  matter field in the adjoint representation of $\tmop{SU} (2)_L$. The dark
  matter candidate corresponds to the neutral component of the new vector
  field ($V^0$). Previous studies have shown that the model can explain the observed dark matter abundance while evading direct and indirect searches. At colliders, the attention has been put on the production of the charged
  companions of the dark matter candidate. In this work, we focus on the mono-Higgs and
  mono-$Z$ signals at hadron colliders. The new charged vectors ($V^{\pm}$) are invisible unless a dedicated search is performed. Consequently, we assume that the mono-Higgs and mono-$Z$ processes correspond {\color{black}to} the $pp\rightarrow h V^{+,0} V^{-,0}$ and $pp\rightarrow Z V^{+,0} V^{-,0}$ reactions, respectively.  We show that, while the $p p \rightarrow h V^{+,0} V^{-,0}$ is more important, both channels may produce significant signals at the HL-LHC and colliders
  running at $\sqrt{s} = 27$ TeV and $100$ TeV, probing almost the complete parameter space.

\end{abstract}

{\maketitle}

\section{Introduction}

The quest for an explanation of dark matter's nature is one of the main
drivers in current High Energy Physics research program. Although more exotic
alternatives exist, the general lore assumes the dark matter (DM) is in fact a
new kind of particle. We already know that such a particle should be massive,
electrically neutral (or with a very small electric charge) and stable (or
with a lifetime of the order of the age of the universe). However, current
data do not significantly constrain its mass nor its spin. Still, an intriguing coincidence
can be noticed: if the DM's mass is of the order of the GeV's, and it interacts
with the standard particles with a strength comparable to weak interaction,
then the observed abundance of DM is naturally reproduced. This is called
``WIMP Miracle'' because it characterizes a ``Weakly Interacting Massive
Particle'' as a well motivated candidate for DM. This observation makes
natural the assumption that the DM feels the weak interaction or, in more
technical words, it should be a neutral component of a $\tmop{SU} (2)_L$
multiplet. This idea has been developed in the so called ``Minimal Dark
Matter'' program {\cite{Cirelli2006}}. Originally, this theoretical setup
was studied only for the cases of particles with spin-$0$ and
spin-$\frac{1}{2}$ in several $\tmop{SU} (2)_L$ representations. However,
there is no fundamental reasons for excluding spin-$1$ particles from this
program. Indeed, the cases of vector particles in the fundamental {\cite{Saez:fundamental}} and adjoint {\cite{vtdm:adjoint}} representations
have been already proposed as viable sources for DM candidates. Typically,
this models must be treated as effective theories, since the presence of
massive vector particle induces perturbative unitarity violation. Nevertheless,
this is not an insurmountable difficulty, especially in the case of the
adjoint representation. In this case, it has been shown that the scale of
unitarity violation is rather high (around $100$ TeV) {\cite{vtdm:adjoint}}. Although an ultraviolet completion of the model has
been constructed {\cite{Abe:2020mph}}, the effective low energy model is
completely adequate for phenomenological purposes. For this reason, in this
paper we focus on the effective theory which extends the Standard Model by the
inclusion of a massive vector field in the adjoint representation (the so
called ``Minimal Vector Dark Matter'' --or MVDM--
model) {\cite{vtdm:adjoint}}. The neutral component of the new massive
vector triplet ($V^0$) becomes the dark matter candidate and it is stabilized
by a $Z_2$ symmetry, which is imposed by reasons related to unitarity preservation {\cite{vtdm:adjoint,Zerwekh:2012bf}}. Previous studies {\cite{vtdm:adjoint}} have shown that the model can explain the observed
dark matter abundance while evading direct and indirect searches{\color{black}, in fact, the expected sensitivity of future experiments such as XENONnT at 20 ton-years {\cite{XENON:2020kmp}} could strongly constrain the parameter space of the studied model.} At
colliders, the attention {\color{black}has} been put on the production of the charged
companions of the dark matter {\color{black} and the mono-jet signal (both processes were studied in \cite{vtdm:adjoint}). Additionally, in reference \cite{Belyaev:2020dissapearing}, it was studied the possibility of the charged companions of the dark matter be long lived particles, producing displaced vertices in ATLAS and CMS}. In this work, we focus on the mono-Higgs and
mono-$Z$ signals at hadron. colliders. {\color{black} The study of these two processes is motivated by the fact  that the MVDM has only two free parameters and, as it will be shown below, mono-Higgs and the mono-$Z$ production can provide complementary information on these parameters}. We show that, while the $p p \rightarrow h V^{+,0} V^{-,0}$ is more important, both channels may produce significant signals at the HL-LHC and colliders
  running at $\sqrt{s} = 27$ TeV and $100$ TeV.  

The rest of the paper is organized in the following way. In section
\ref{Sec:Model} we briefly describe the main aspects of the MVDM model. Then in section \ref{Sec:Sim_consider} we explain our methodology. In section \ref{Sec:Results} we discuss our results, while in section
\ref{Sec:Conclusions} we state our conclusions. \ \ \ \ \ \

\

\section{The Model}\label{Sec:Model}

For the reader's convenience, we briefly recall the main characteristics of
the model which was fully developed in Ref. \cite{vtdm:adjoint} . The Standard Model (SM) is extended by the introduction of a new triplet vector
field ($V_{\mu} = (V_{\mu}^+, V_{\mu}^0, V_{\mu}^-)^T $) transforming
homogeneously in the adjoint representations of $\tmop{SU} (2)_L$ with no
hypercharge, that is,
\begin{equation}
  V_{\mu} \rightarrow U^{\dag}_L V_{\mu} U_L
\end{equation}
with $U_L \in \tmop{SU} (2)_L$. Additionally,  a $Z_2$ symmetry is imposed, under
which all the standard fields are even and $V_{\mu}$ is odd. The model is,
then, defined by the following Lagrangian:
\begin{eqnarray}
  \mathcal{L} & = & \mathcal{L}_{SM} - \tmop{Tr} \{ D_{\mu} V_{\nu} D^{\mu}
  V^{\nu} \} + \tmop{Tr} \{ D_{\mu} V_{\nu} D^{\nu} V^{\mu} \} \nonumber\\
  &  & - \frac{g^2}{2} \tmop{Tr} \{ [V_{\mu}, V_{\nu}] [V^{\mu}, V^{\nu}] \}
  \\
  &  & - ig \tmop{Tr} \{ W_{\mu \nu} [V^{\mu}, V^{\nu}] \} + \tilde{M}^2
  \tmop{Tr} \{V_{\nu} V^{\nu} \} \nonumber\\
  &  & + a (\Phi^{\dagger} \Phi) \tmop{Tr} \{V_{\nu} V^{\nu} \} \nonumber
\end{eqnarray}
where the covariant derivative $D_{\mu} = \partial_{\mu} - i g [W_{\mu},
\cdot]$ is the usual one for the adjoint representation of $\tmop{SU} (2)_L$.
After the electroweak symmetry breaking, the physical mass of the
new vector triplet is:
\begin{equation}
  M_V^2 = \tilde{M}^2 + \frac{1}{2} a v^2
\end{equation}
where $v$ is the vacuum expectation value of the Higgs field $\langle \Phi
\rangle = v$. 

We want to emphasize that this construction is rather minimal, in the sense that only two new free parameters are added: the mass of the vector triplet $M_V$ and the Higgs-Vector coupling constant $a$.

\begin{figure}[h!]
    \centering
    \includegraphics[scale=0.3]{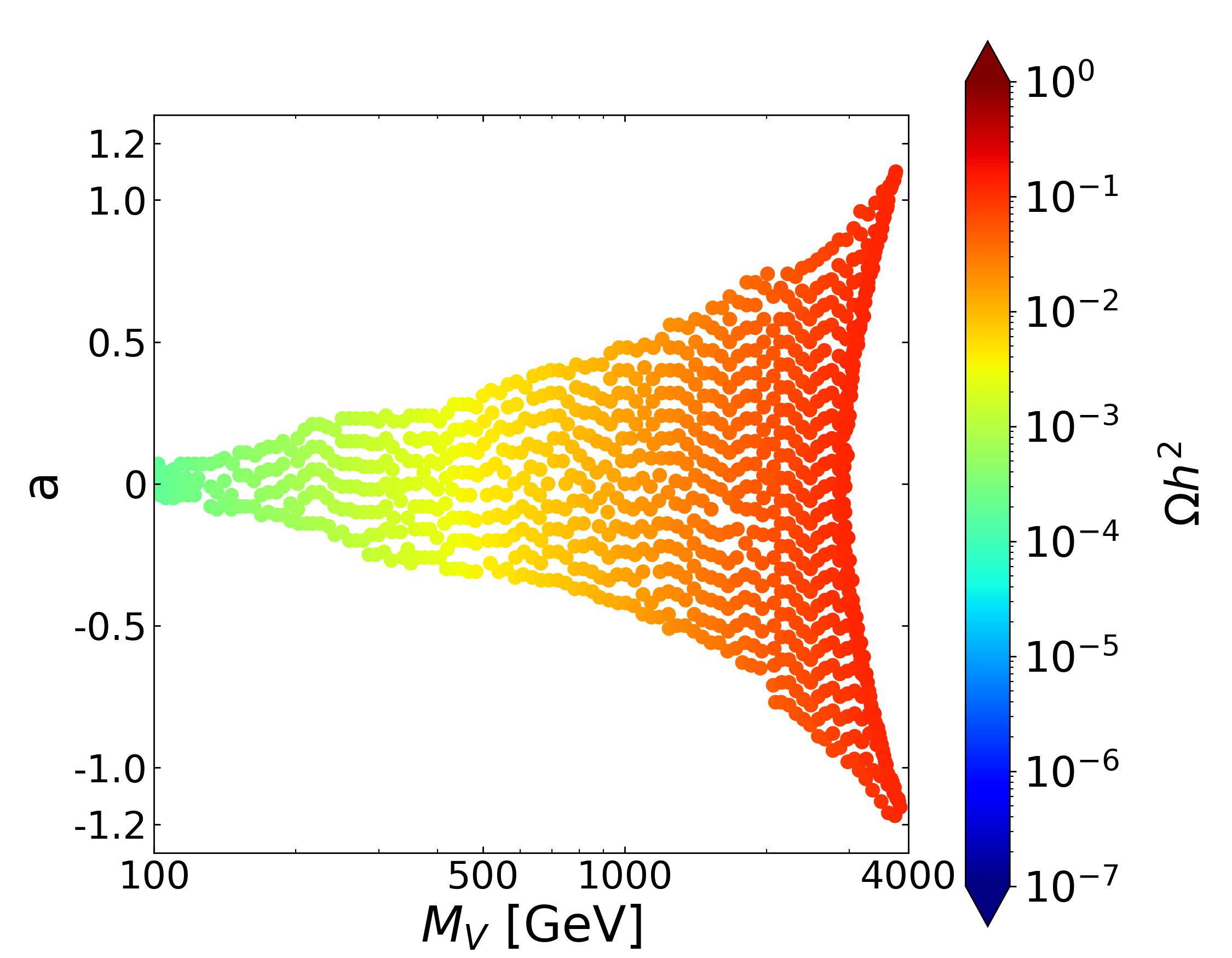}
    \caption{Parameter space allowed by experimental and observational constrains. This includes: direct and indirect dark matter searches, dark matter abundance, consistency with perturbative unitarity. The color scale indicates the value of the relic density. Notice that we have include the case of subabundance.}
    \label{parameter_space}
\end{figure}

At this point, it is worth to recall that radiative {\color{black} electroweak} corrections {\color{black} at loop level}
induce a mass splitting between the charged components of $V_{\mu}$ and the
neutral component. {\color{black} Concretely speaking, the charged components of $V_{\mu}$ receive an additional electromagnetic correction to their propagator, which is absent for the case of $V^0$. For
large values of $M_V$, the mass splitting is given by:}
%This mass splitting is of electromagnetic origin and, for
%large values of $M_V$, its value is:
\begin{equation}
  \Delta M \equiv M_{V^{\pm}} - M_{V^0} \approx 200 \tmop{MeV},
\end{equation}
making $V^0$ the lightest particle which is odd under the $Z_2$ symmetry. Consequently, $V^0$ happens to be stable and a good DM candidate. Indeed, the model reproduces the
observed relic density \cite{Planck:2018vyg} while remaining consistent with all the experimental
constraints if $M_V$ in the range from 3 to 4 TeV. In Figure \ref{parameter_space}, we show the region of the parameter space which is allowed by current experimental constrains.

As a consequence of the tiny mass splitting, the only decay channel available
to the charged components of the vector triplet is $V^{\pm} \rightarrow V^0 \pi^{\pm}$ where the pion happens to be very soft.
Consequently, it is very difficult to detect any $V^{\pm}$ except by dedicated searches for long living particles \cite{Belyaev:2020dissapearing,Fukuda:2017disstrackcharged,Saito:2019disstrackcharged,Mahbubani:2017disstrackcharged}. As a result, we assume that $V^0$ and $V^{\pm}$
contribute to the missing energy  when they are present in the final state.

\section{Methodology}\label{Sec:Sim_consider}

We start by considering the mono-Higgs signal associated to the $V^{\pm,0}$ production. Our main and irreducible background is the \textit{Higgs-Strahlung} with a $Z$-boson decaying into neutrinos \cite{Belyaev:2020dissapearing}. Although there are other sources of background such as $t\Bar{t}$ and $Zb\Bar{b}$ processes, they can be filtered out by applying appropriate kinematic cuts, implementing a lepton veto, and enhancing the trigger in future experiments \cite{Ghorbani:2016edw,ATLAS:2021shl,No:2015xqa,Carpenter:2013xra,Bhowmik:2020spw}, consequently they will not be taken into account in this analysis.

We used the \verb|CalcHEP| \cite{calchep} package to perform our simulations. The signal was computed using the \verb|CalcHEP| implementation of the model \cite{vtdmp} available at the High Energy Physics Model Database \cite{hepmdb}.

The main process contributing to the signal of $V^0$ production, is $gg\to hV^0V^0$. Since the $V^0$ does not couple to fermions and a $Z V^0 V^0$ vertex does not exist, the only way to have a $q\bar{q} \to hV^0V^0$ process is through the direct coupling of the Higgs boson to quarks, making it highly suppressed. 

Additionally, we took into account the Higgs boson decay into two $b$-quarks. The reconstruction efficiency for this channel is estimated to be $\epsilon_{eff-b}\approx 0.80$ \cite{CMS:2012feb_brecosntruction,ATLAS:2021breconstruction,ATLAS:2017breconstruction,ATLAS:2015breconstruction} and the $h\to b\Bar{b}$ decay probability \cite{higgs_resume_2022}, $Br(h\to b\Bar{b}) \approx 0.58$ . Using the standard definition of statistical significance as a function of the integrated luminosity on the collider:

\begin{equation}
    S=\sqrt{\epsilon_{eff-b}\cdot Br(h\to b\Bar{b}) \cdot L} \cdot \dfrac{\sigma_{
    s}}{\sqrt{\sigma_{
    s}+\sigma_{b}}}
    \label{statistical_significance_higgs}
\end{equation}

where $\sigma_s=\sigma(pp\rightarrow h V^{+,0} V^{-,0})$ and $\sigma_b=\sigma(pp\rightarrow Z h \rightarrow \nu \Bar{\nu} h)$ are the cross-section for the signal and the background respectively, extracted from the simulations made on \verb|CalcHEP| and $L$ the integrated luminosity.

On the other hand, the main contribution to the background for Mono-$Z$ production comes from the $pp\to ZZ$ channel. In this case, we have to impose that one of the $Z$-bosons decays into visible particles while the other one, into  neutrinos. Consequently, for the mono-$Z$ process, our significance formula can be written as:  

\begin{equation}
    S=\sqrt{Br(Z\to visible) \cdot L} \cdot \dfrac{\sigma_{
    s}}{\sqrt{\sigma_{
    s}+\sigma_{b}}}
    \label{statistical_significance_z}
\end{equation}

where this time $\sigma_s=\sigma(pp\rightarrow Z V^{+,0} V^{-,0})$ and $\sigma_b=\sigma(pp\rightarrow Z Z \rightarrow \nu \Bar{\nu} Z)$. {\color{black} In our discussion, we will pay attention to the regions of the parameter space where we get $S>2$. Of course, a low significance like $S=2$ indicates only a weak hint of the possible existence of New Physics and only regions with $S>5$ provide the conditions needed for a discovery.}
\\
\\

\begin{figure}[h!]
    \centering
    \subfloat[{\color{black}$pp\to hVV$}]{\label{mono-h_diagram}
    \includegraphics[width=0.588\linewidth]{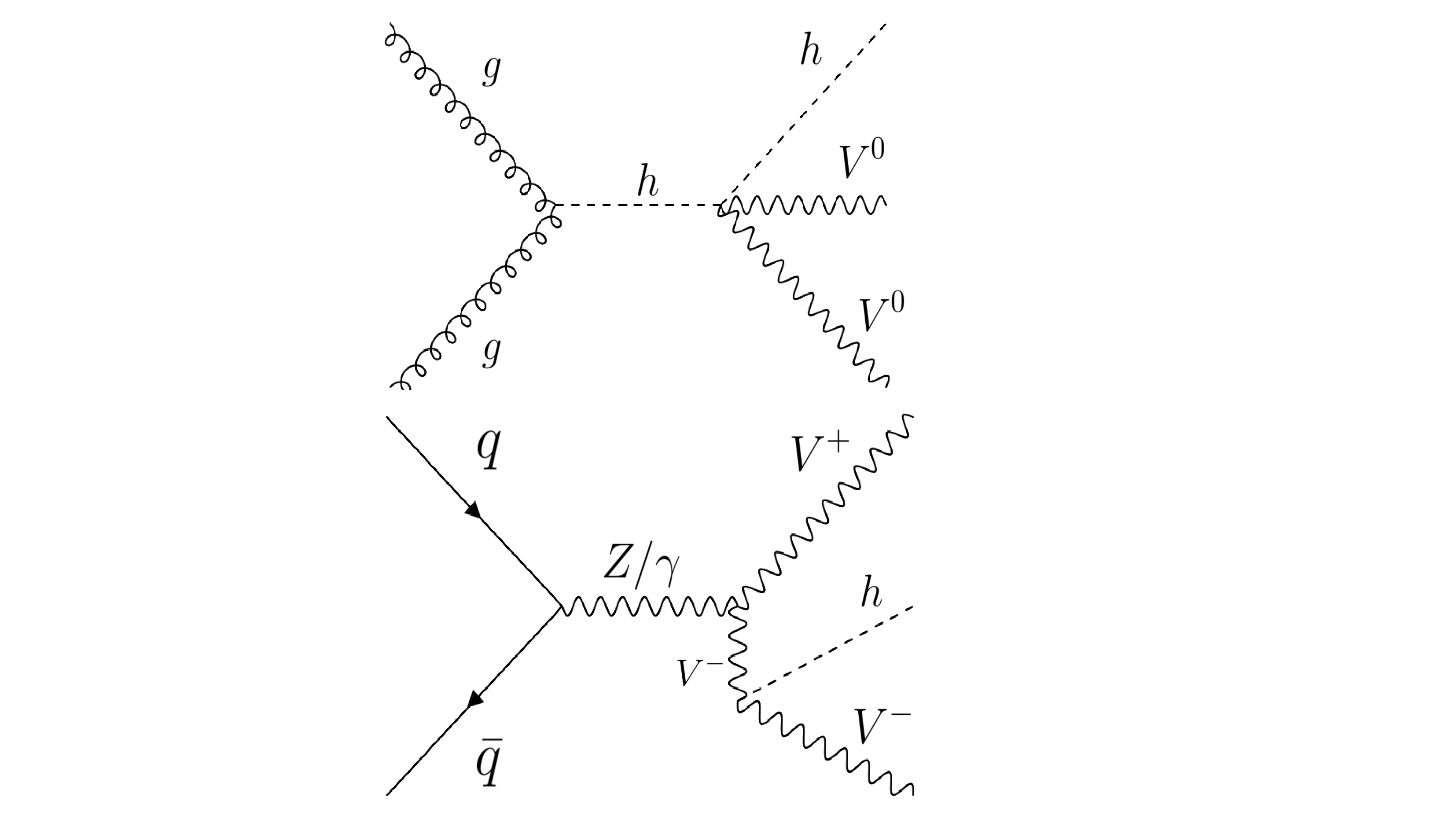}}
    \hspace{-0.2\linewidth}
    \subfloat[{\color{black}$pp\to ZVV$}]{\label{mono-z_diagram}
    \includegraphics[width=0.588\linewidth]{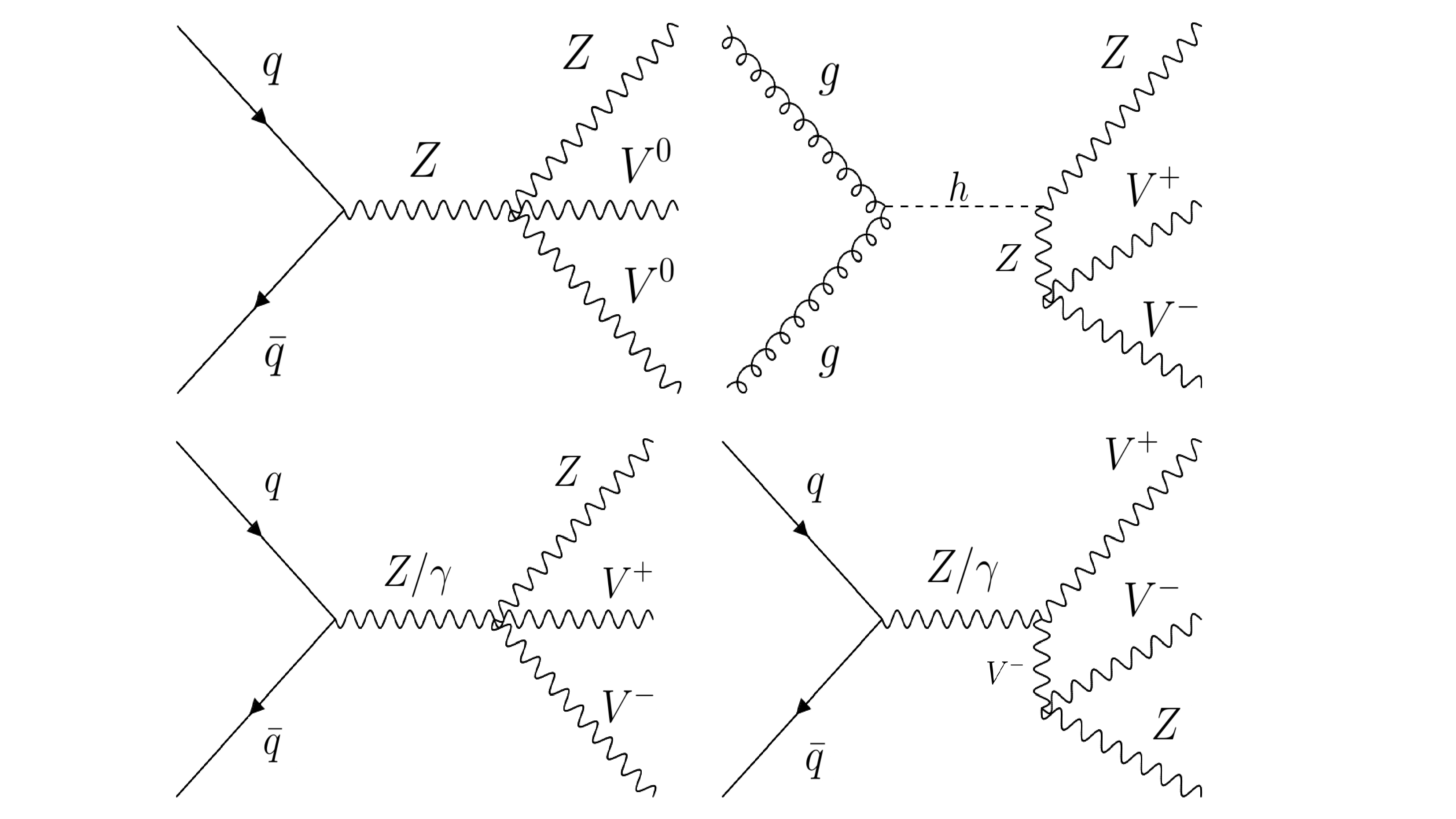}}
    \caption{{\color{black}Representative diagrams of the signal process used in the analysis. The diagrams include the effective $hgg$ vertex generated by a loop of top quarks. }}
    \label{signals}
\end{figure}

At this point, it is valuable to remark that the cross section for the process $pp\rightarrow X V^{+} V^{-}$ ($X=h,Z$) is in general much bigger than $\sigma(pp\rightarrow X V^{0} V^{0})$ due to the fact the first one contains electroweak contributions that the last one lacks {\color{black}(see the production diagrams in Figure \ref{signals})}.

We performed simulations for three future facilities: the HL-LHC, the HE-LHC and the FCC-hh. Their projected energies and its expected maximum integrated luminosities are showed in Table \ref{table_cm_luminosity_mono-X}. 

\begin{table}[htb]
    \centering
    \setlength{\tabcolsep}{10pt}
    \begin{tabular}{|c||c|c|c|}
        \hline
        Collider&HL-LHC&HE-LHC&FCC-hh\\
        \hline
        &&&\\
        $\sqrt{s}$   \,[TeV]&13.6&27&100 \\
        &&&\\
        \hline
        \hline
        $L\,\,[ab^{-1}]$&3 &10 &30 \\
        \hline
    \end{tabular}
    \caption{Integrated Luminosities and energies used in our simulations. The values were taken from \cite{Cepeda:2019klc,FCC:2018vvp}}
    \label{table_cm_luminosity_mono-X}
\end{table}

\begin{figure}[h!]
    \centering
    \includegraphics[width=0.425\linewidth]{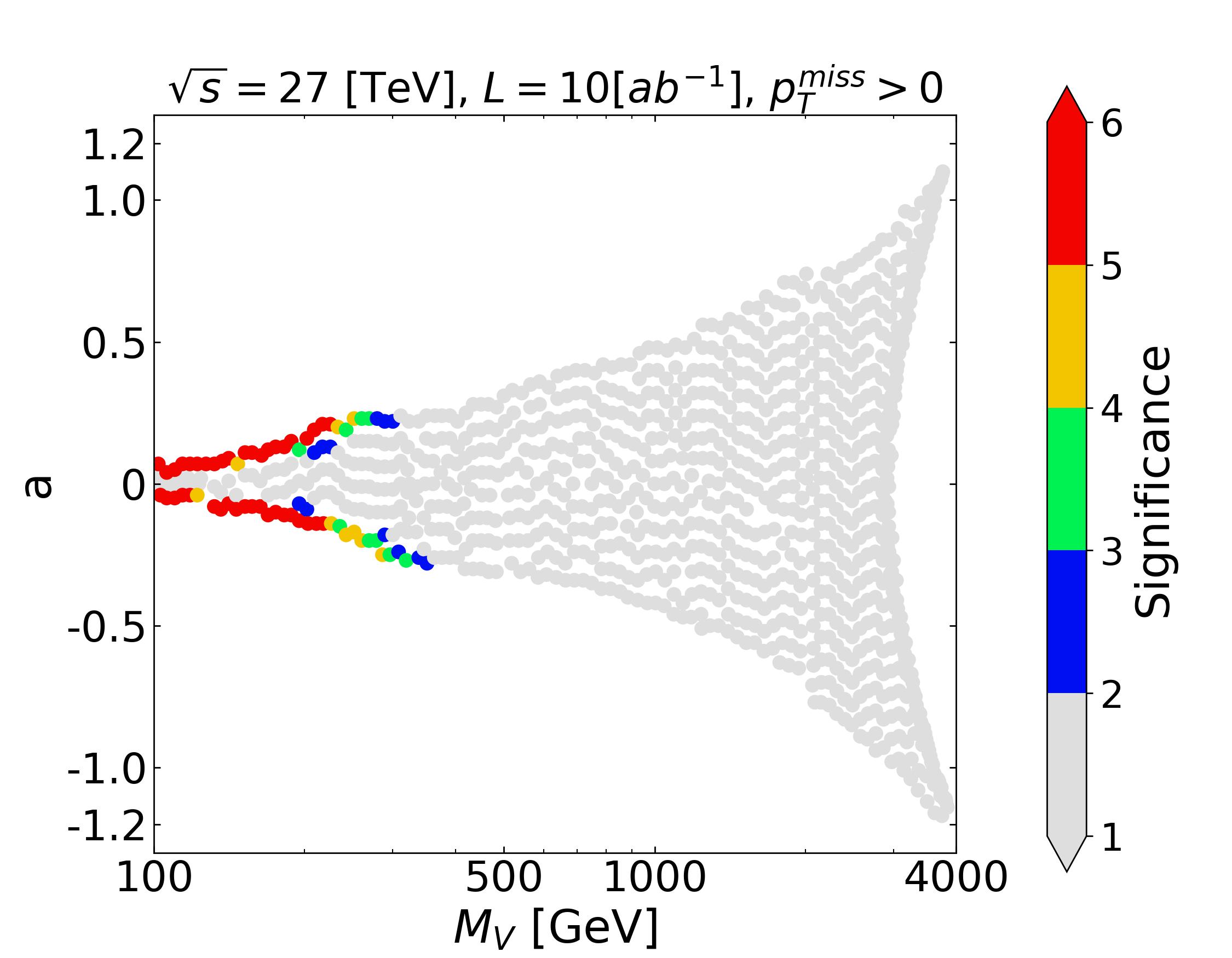}
    \includegraphics[width=0.425\linewidth]{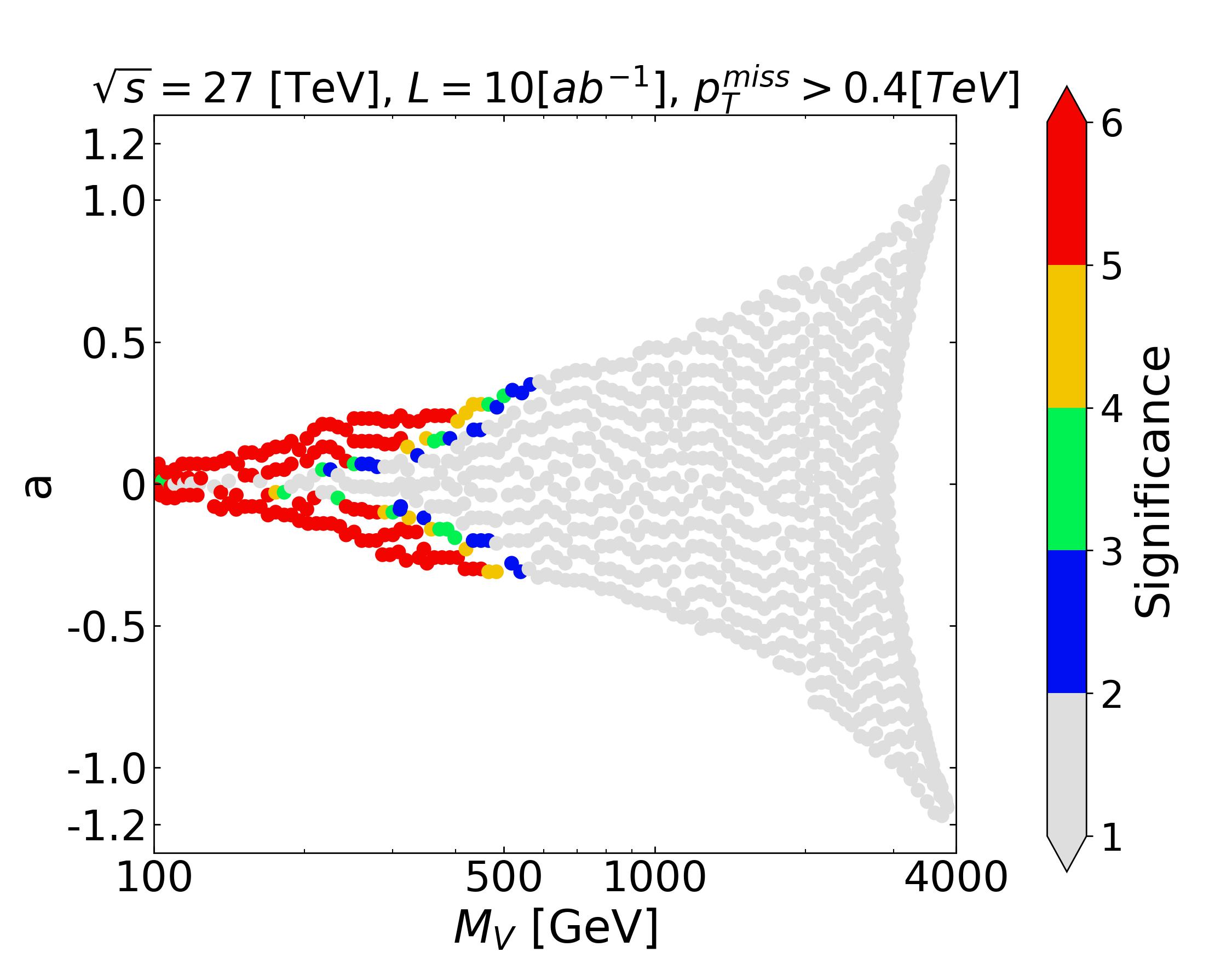}
    \includegraphics[width=0.425\linewidth]{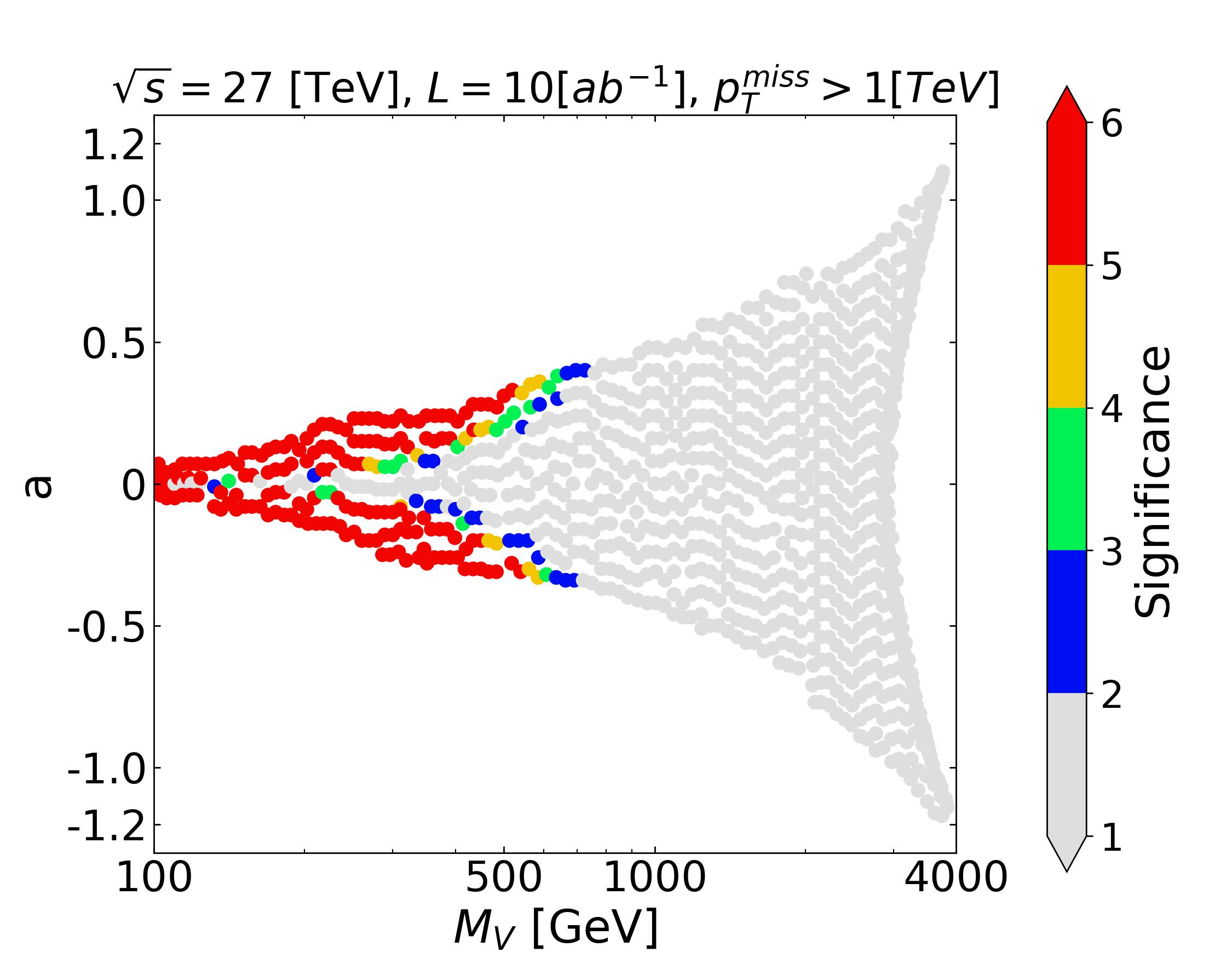}
    \includegraphics[width=0.425\linewidth]{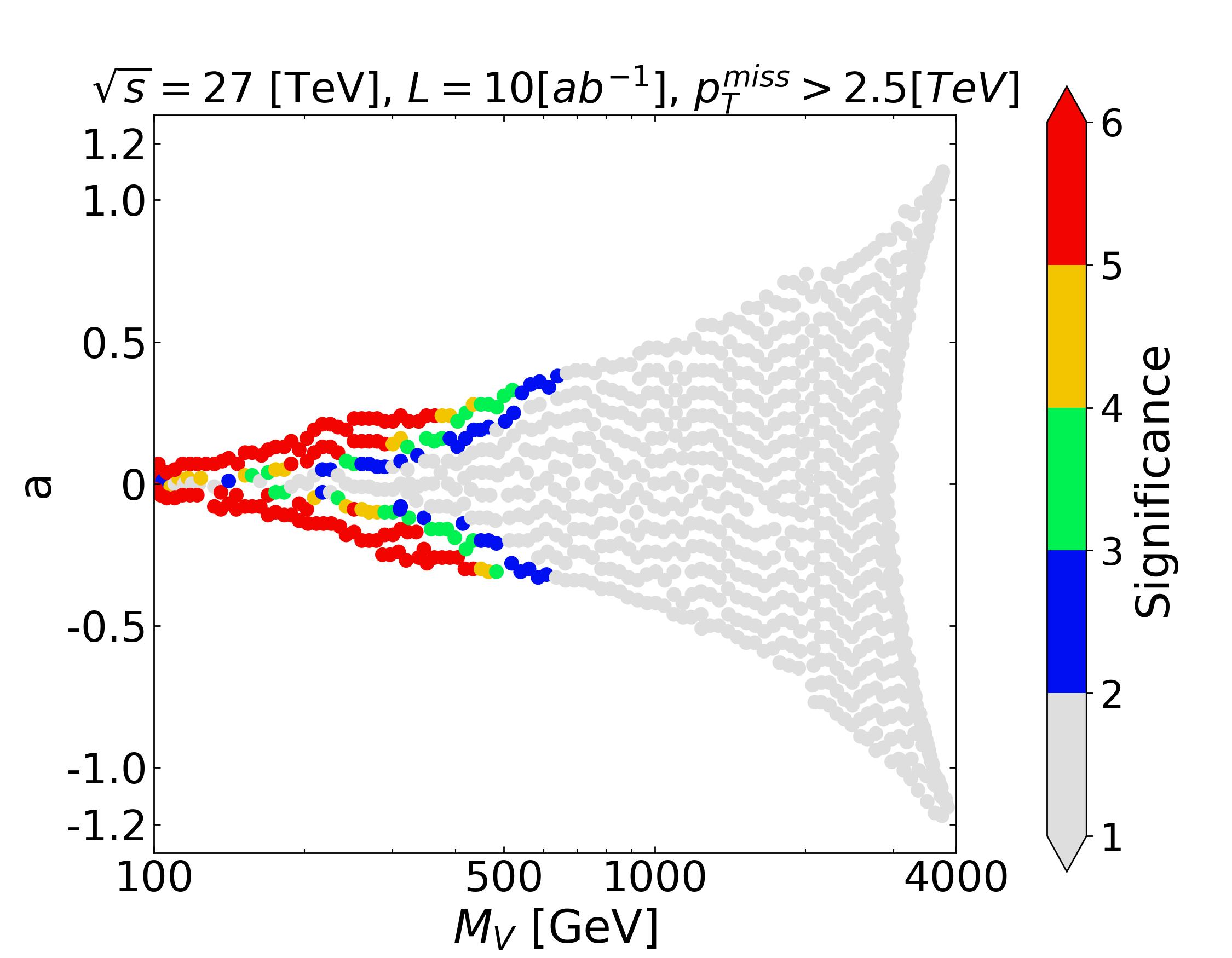}
    \caption{Parameter space displaying the statistical significance and satisfying the condition $S>2$, for the HE-LHC with an integrated luminosity $L=10\,ab^{-1}$ under various kinematic cuts in the mono-Higgs event. }
    \label{best_cut}
\end{figure}

An important aspect of our simulations was to set up kinematic cuts in order to optimize the statistical significance of the signal. We focused on the missing transverse momentum ($p_T^{\text{miss}}$, {\color{black} which corresponds to the apparently unbalanced transverse momentum of the $h$ or $Z$ bosons}) as our relevant kinematic variable. We explored the parameter space using different cuts in $p_T^{\text{miss}}$ until we obtained an optimal cut. An example of this process is shown in Figure \ref{best_cut}.

The optimal kinematic cut results for mono-Higgs and mono-$Z$ events across the three accelerators are shown in Table \ref{ecm_corte_optimo}.

In all our simulations, we employed  the \verb|cteq6l(proton)| parton distribution function {\color{black} \cite{cteq6l}}. Finally, we will base our work in the final parameter space found in reference \cite{vtdm:adjoint}  and represented in Fig \ref{parameter_space}.

\begin{table}[htbp]
    \centering
    \setlength{\tabcolsep}{10pt}
    \begin{tabular}{|c||c|c|c|}
        \hline
        Collider&HL-LHC&HE-LHC&FCC-hh\\
        %&&&\\
         %$\sqrt{s}$ [TeV]& 13.6 & 27 & 100 \\
         %&&&\\
         \hline
         \hline
         &&&\\
         {\color{black}$p_T^{\text{miss}}$} [TeV]& 0.4 & 1 &3.5 \\
         &&&\\
         \hline
    \end{tabular}
    \caption{Optimal Kinematic cuts for the missing momentum transverse of the Higgs and $Z$-boson, according to the collider.}
    \label{ecm_corte_optimo}
\end{table}

\clearpage
\newpage

\section{Results}\label{Sec:Results}

\subsection{Mono-Higgs Production}\label{subsec:Mono-Higgs_production}

We start by considering the mono-Higgs production. As explained above, this process {\color{black}receives} contributions not only from the associated production of a Higgs boson and two $V^0$'s, but also from the production of a Higgs with a pair of charged massive vectors. In order to visualize the relative importance of each subprocess, we will discuss first the $h V^0 V^0$ channel and later we will consider the mono-Higgs production taken into account all the contributions.

\subsubsection{$V^0$ production}\label{subsubsec:Mono-Higgs_V0}

\begin{figure}[htbp]
    \centering
    \includegraphics[width=0.373\linewidth]{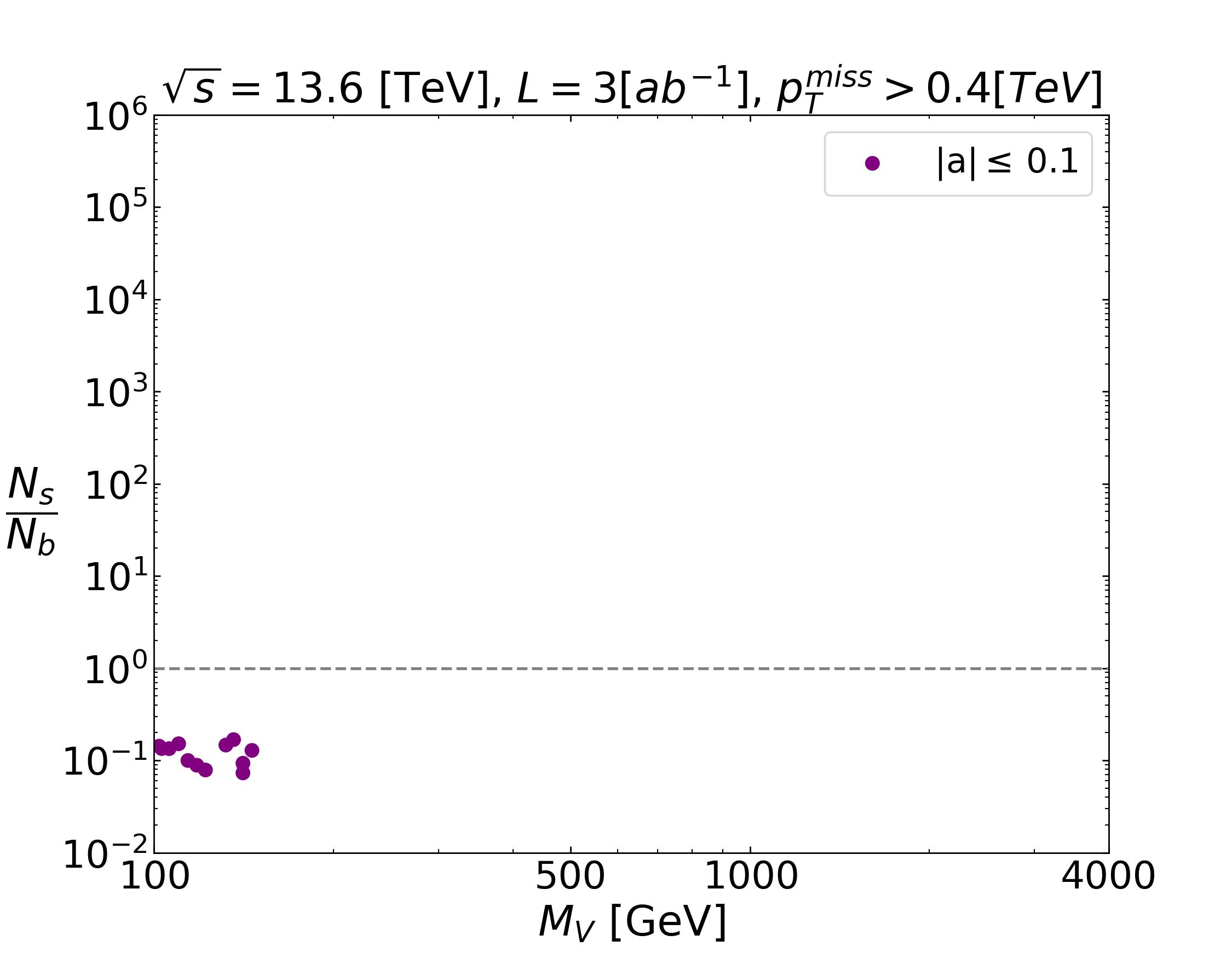}
    \includegraphics[width=0.373\linewidth]{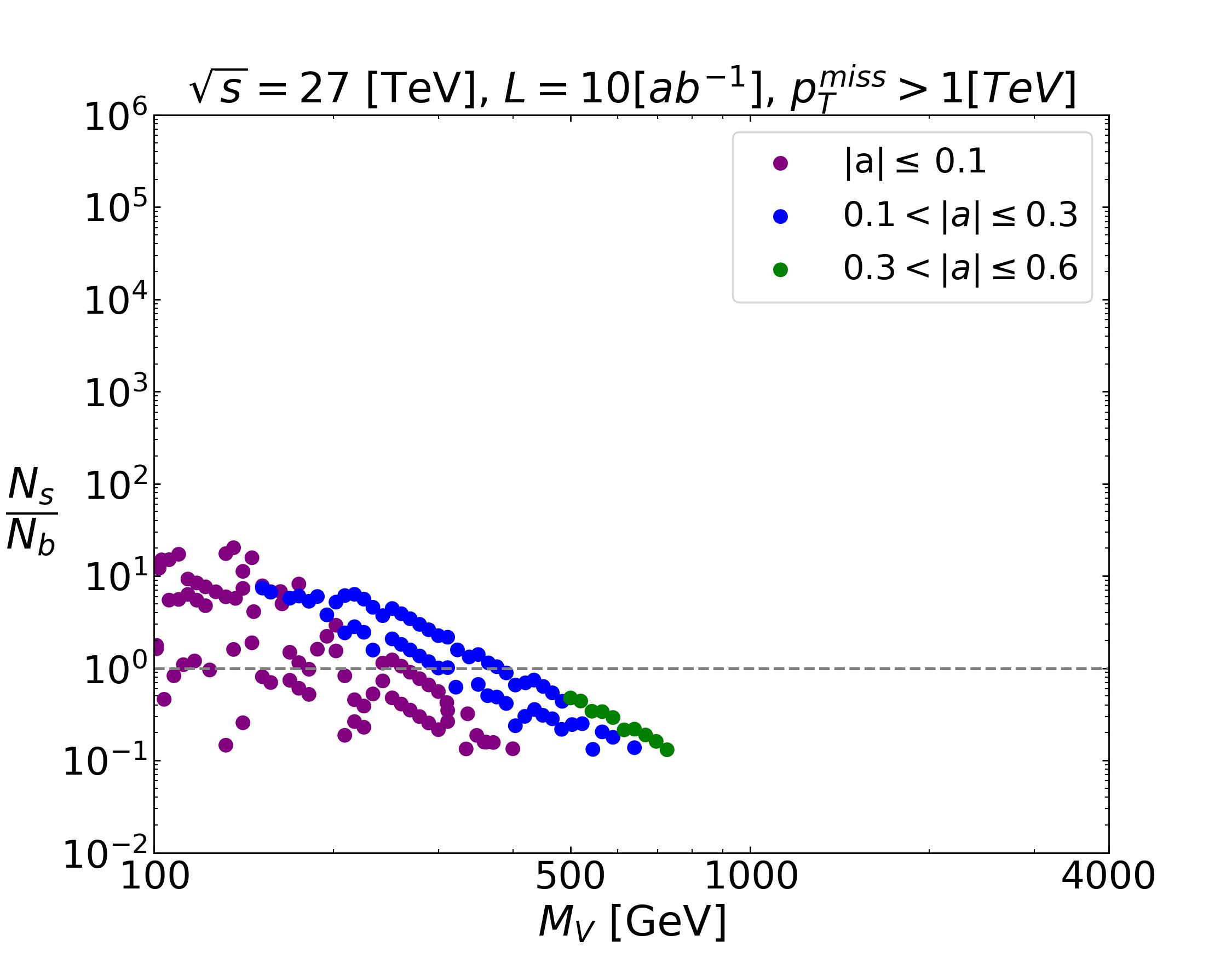}
    \includegraphics[width=0.373\linewidth]{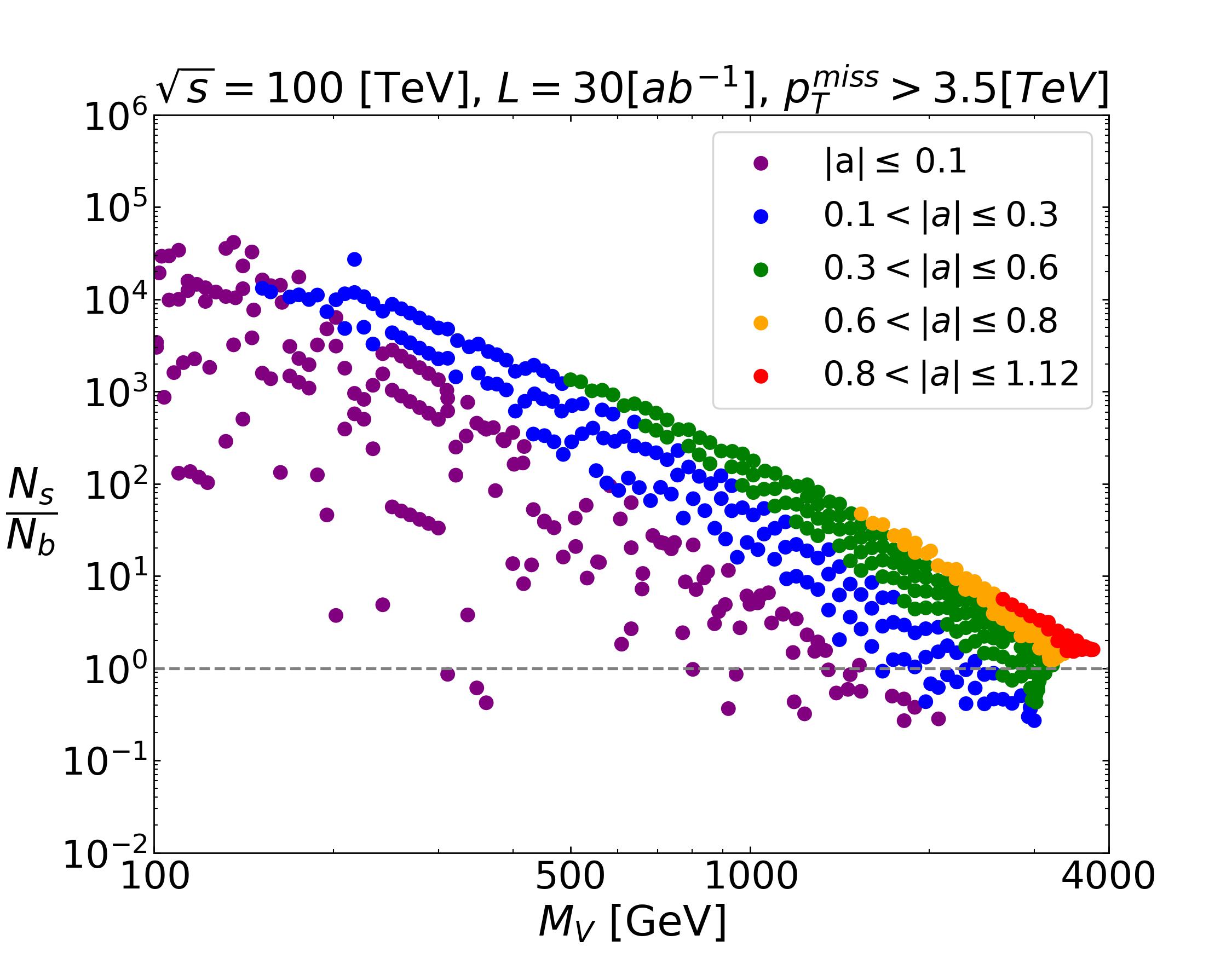}

    \caption{Ratio between signal and background events for each accelerator {\color{black}for the process} $pp\to hV^0V^0$. The {\color{black}dashed gray lines} indicates when the ratio is 1. All the points satisfy the constrain {\color{black}$S>2$.} }
    \label{events_excess_hV0V0}
\end{figure}

\begin{figure}[htbp]
    \centering
    \includegraphics[width=0.373\linewidth]{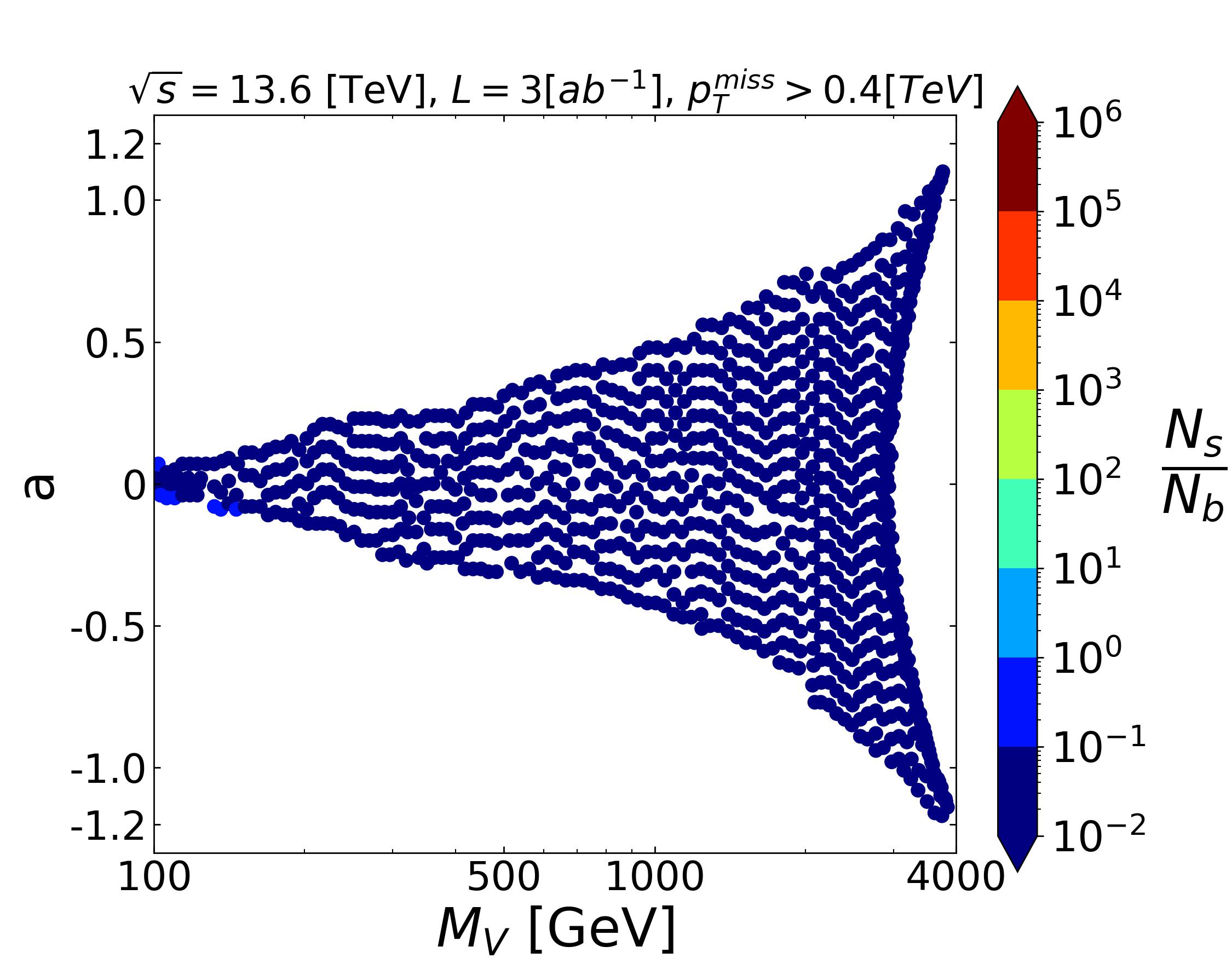}
    \includegraphics[width=0.373\linewidth]{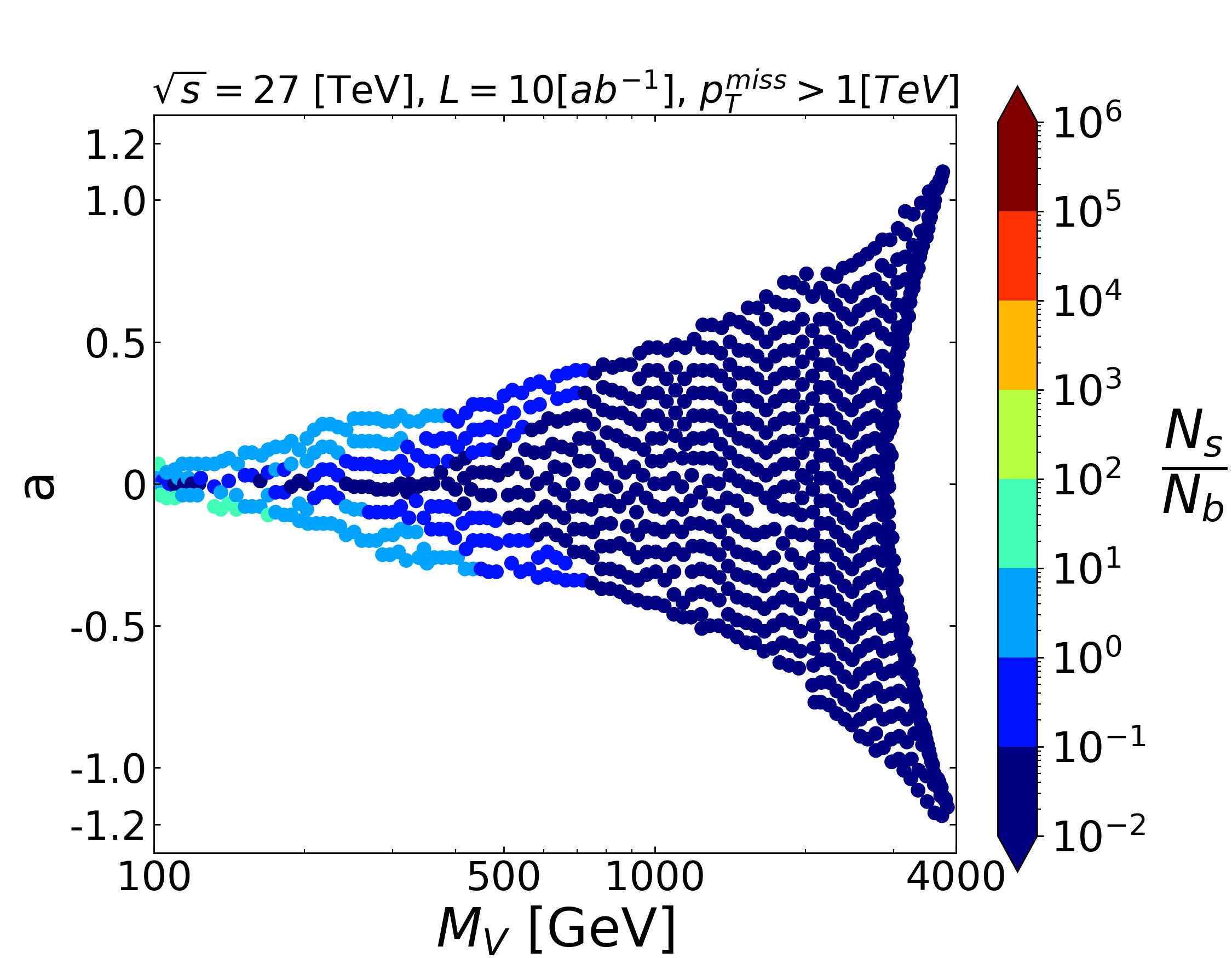}
    \includegraphics[width=0.373\linewidth]{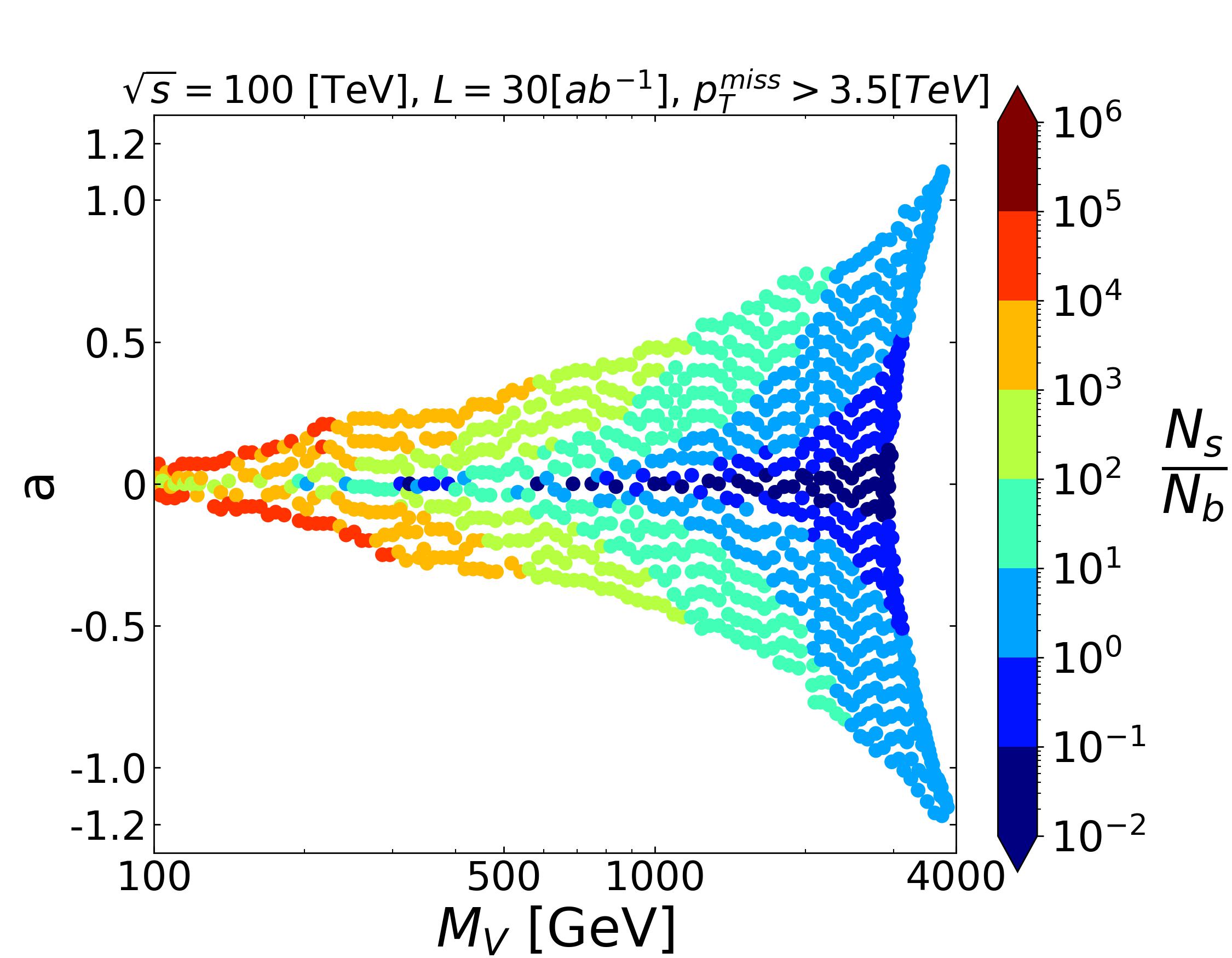}
    \caption{Parameter space displaying the value of the ratio between signal and background events in the color scale, for each accelerator {\color{black}for the process} $pp\to hV^0V^0$.}
    \label{events_excess_hV0V0_avsMv}
\end{figure}

To estimate the observability potential of our model in the $h V^0 V^0$ channel, we computed the signal to background ratio and the statistical significance of the signal as functions of the DM mass ($M_V$) and the Higgs-DM coupling constant ($a$). 
In Figure \ref{events_excess_hV0V0} we show the signal to background {\color{black}ratio} as a function of $M_V$ under the condition that the statistical significance be larger than 2. Meanwhile in Figure \ref{events_excess_hV0V0_avsMv} we show which region of the parameter space is tested with the same criterion. 

\begin{figure}[htbp]
    \centering
    \includegraphics[width=0.373\linewidth]{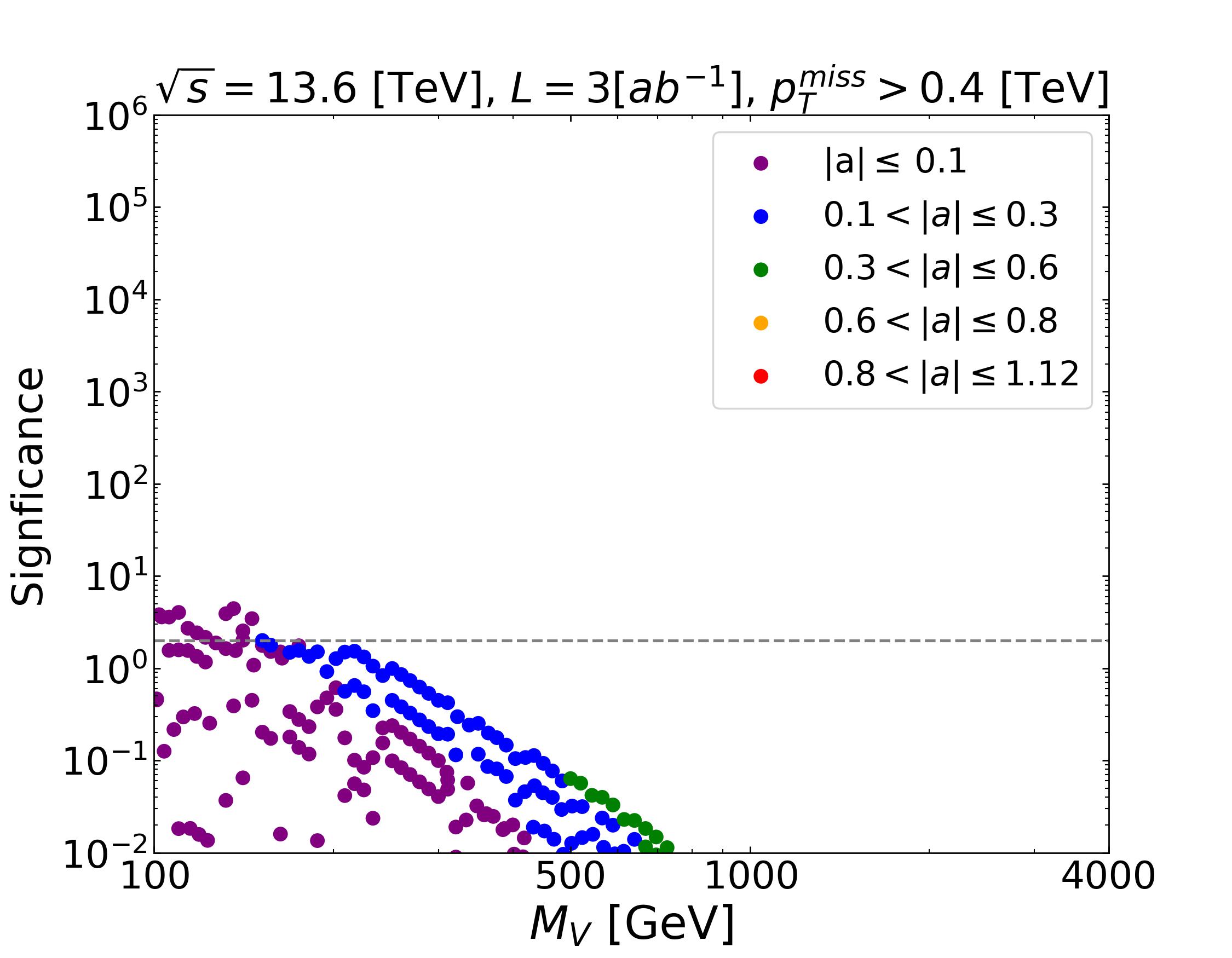}
    \includegraphics[width=0.373\linewidth]{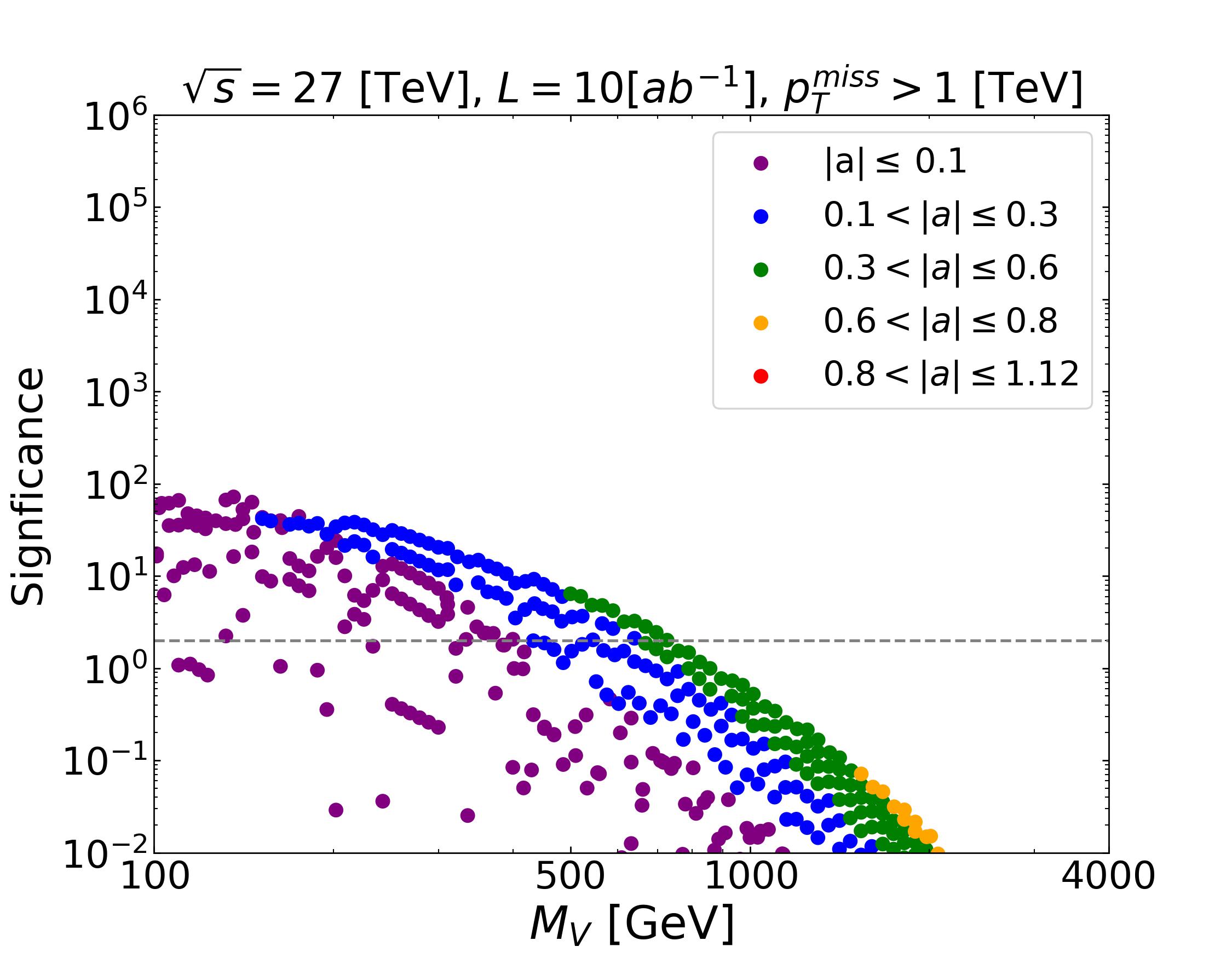}
    \includegraphics[width=0.373\linewidth]{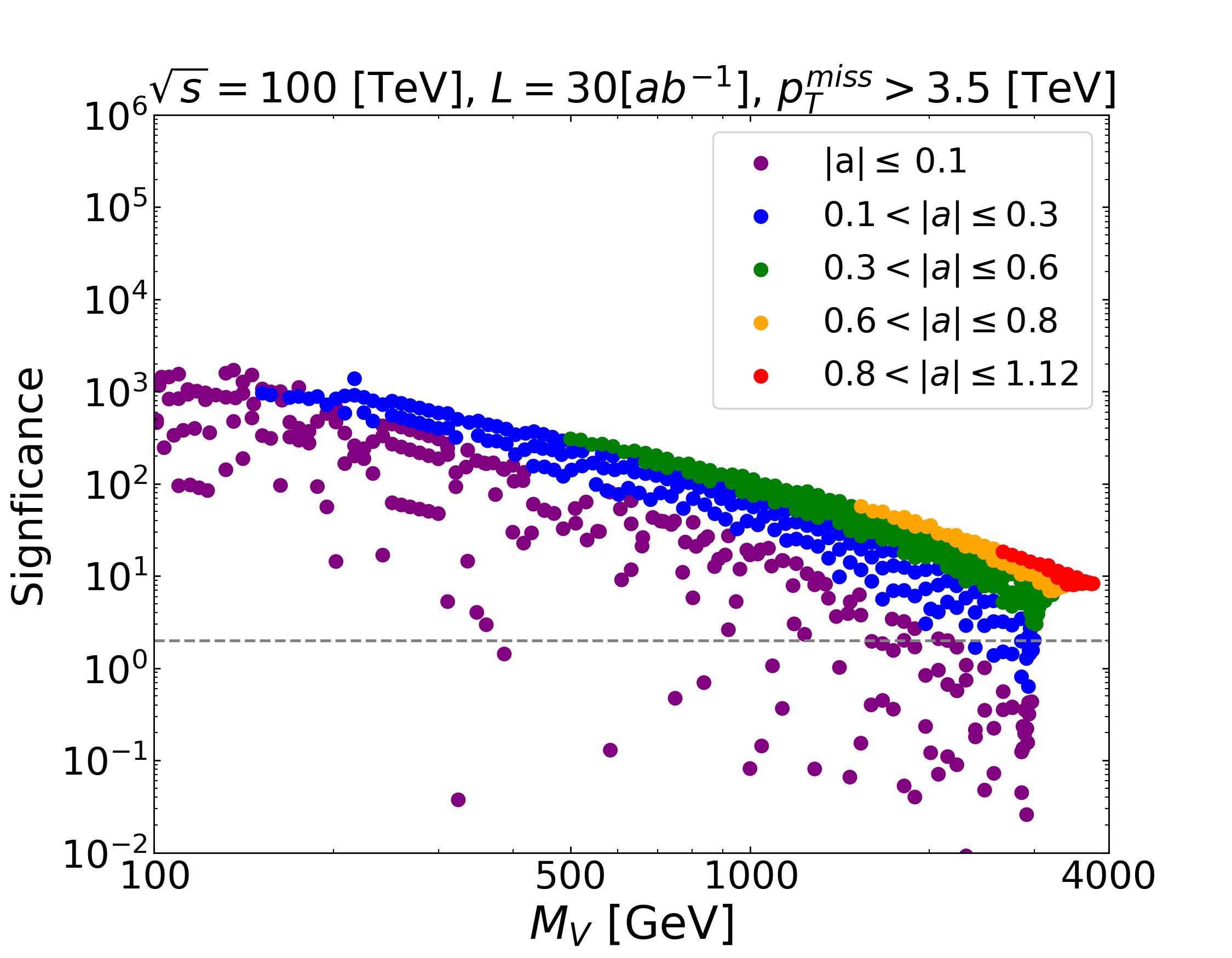}
    \caption{Statistical significance as function of the vector mass {\color{black}for the process} $pp\to hV^0V^0$. The {\color{black}dashed gray lines} indicates when the significance reach the 68\% confidence level ($S=2$).}
    \label{significance_hv0v0_sigmavsMv}
\end{figure}

\begin{figure}[htbp]
    \centering
    \includegraphics[width=0.373\linewidth]{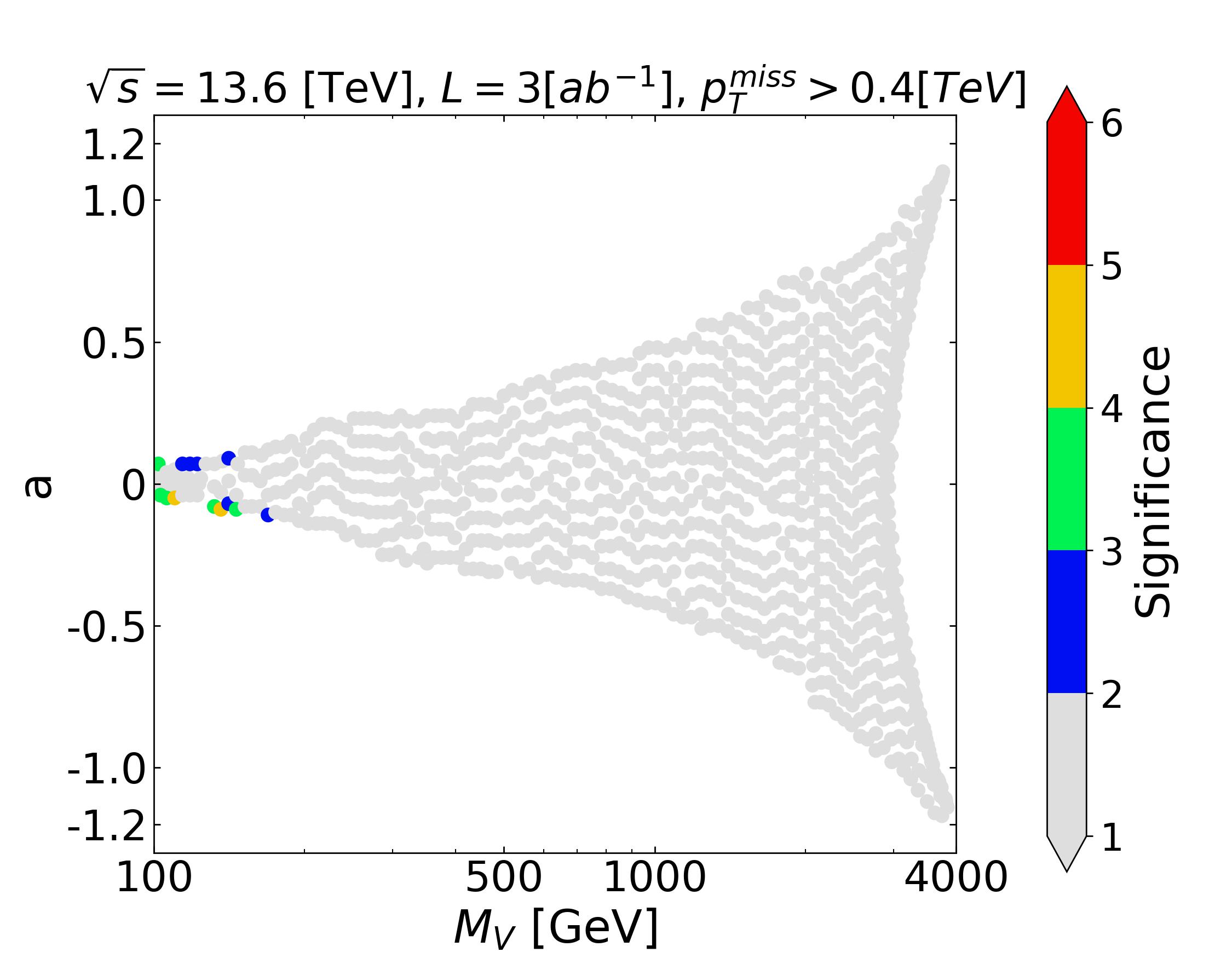}
    \includegraphics[width=0.373\linewidth]{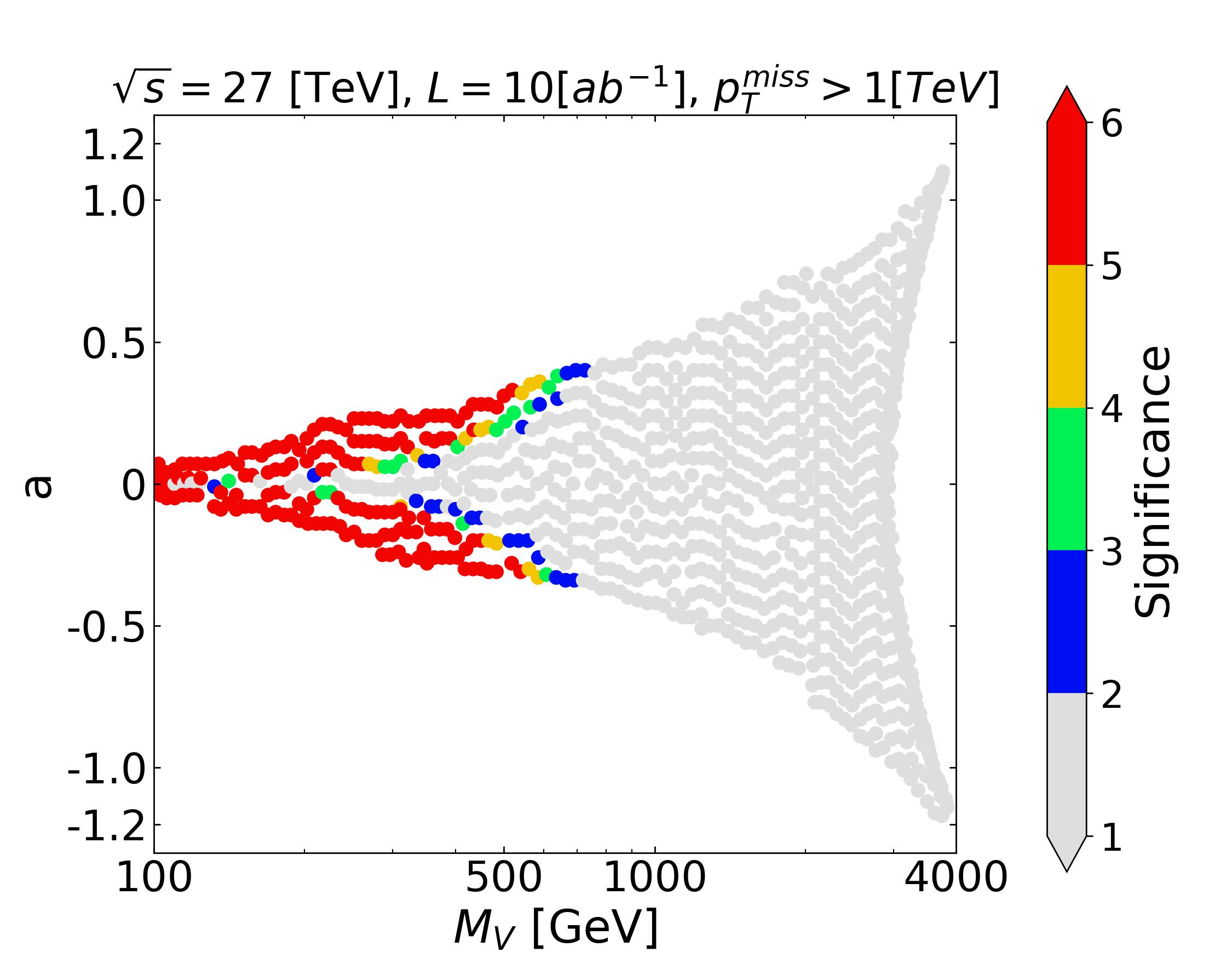}
    \includegraphics[width=0.373\linewidth]{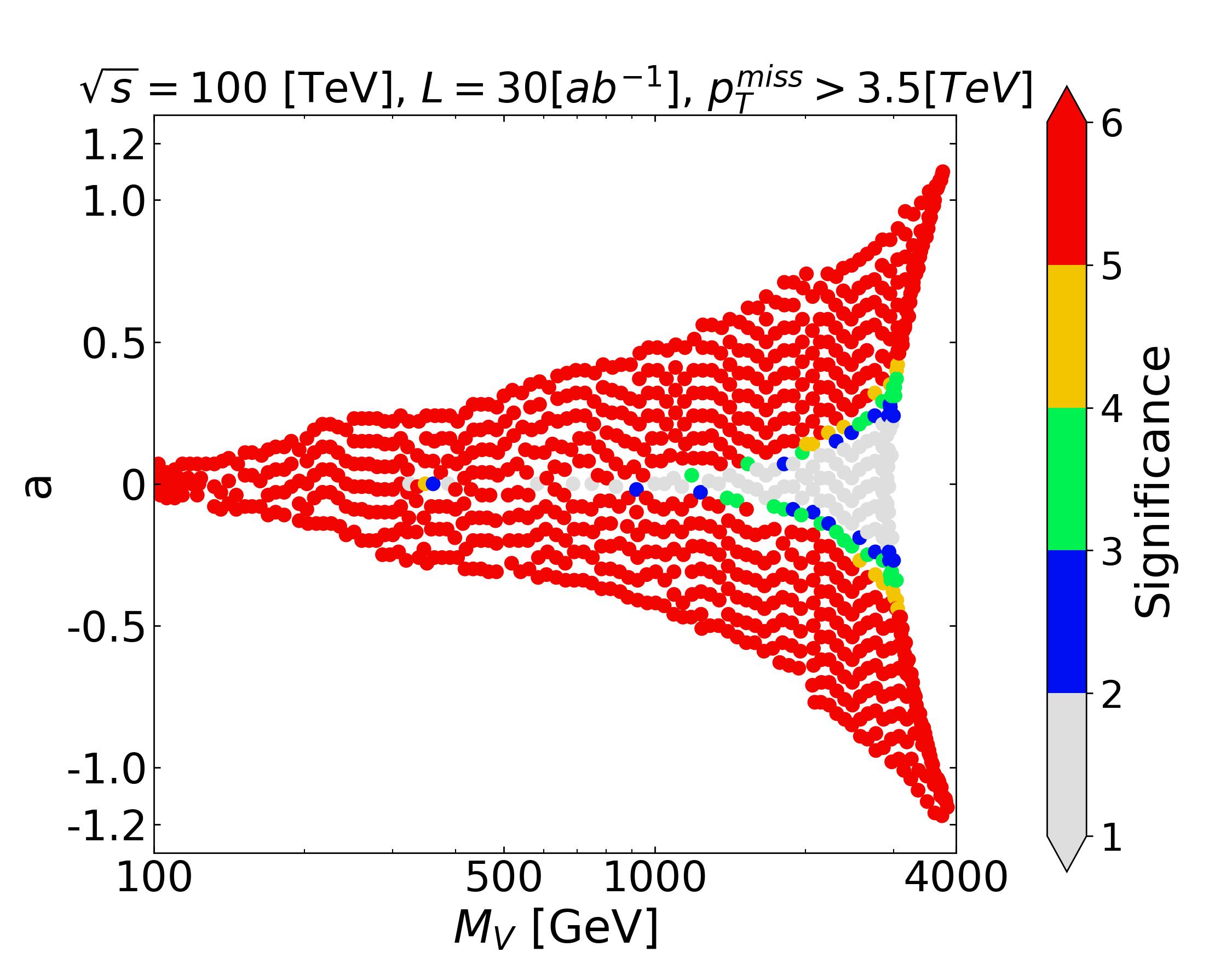}
    \caption{Parameter space showing the Significance in the color bar {\color{black}for the process} $pp\to hV^0V^0$ @ 13.6; 27 and 100 TeV of energy center of mass with its respective optimal kinematic cut and maximum integrated luminosity.}
    \label{significance_hv0v0_avsMv}
\end{figure}

Figures \ref{significance_hv0v0_sigmavsMv} and \ref{significance_hv0v0_avsMv} are similar to the previous ones, but this time the statistical significance is the relevant quantity. We see that the HL-LHC cannot prove the model {\color{black} (reaching S $\simeq$ 4)}, the FCC has the potential to test almost the whole parameter space except for a small region of high DM mass and small Higgs-DM coupling. In fact, if the $V^0$ were the only source of missing energy, the mono-Higgs production could provide relevant signal of the model for $M_V\lesssim$ {\color{black}$550$}  GeV at the HE-LHC and $M_V\lesssim 4$ TeV for the FCC.

\clearpage
\newpage

\subsubsection{$V^{\pm,0}$ production}\label{subsubsec:Mono-Higgs_VpVm}

\begin{figure}[htbp]
    \centering
    \includegraphics[width=0.373\linewidth]{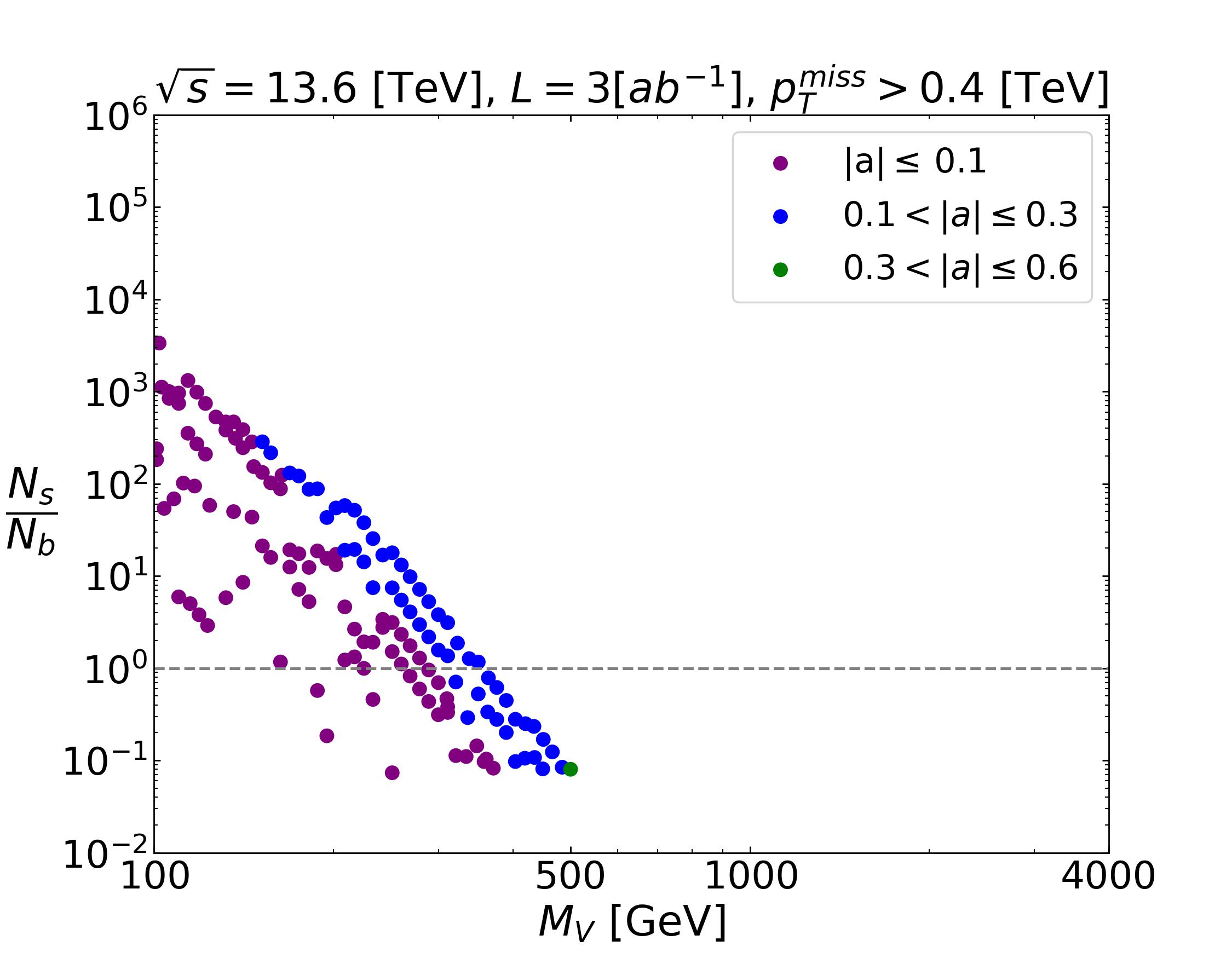}
    \includegraphics[width=0.373\linewidth]{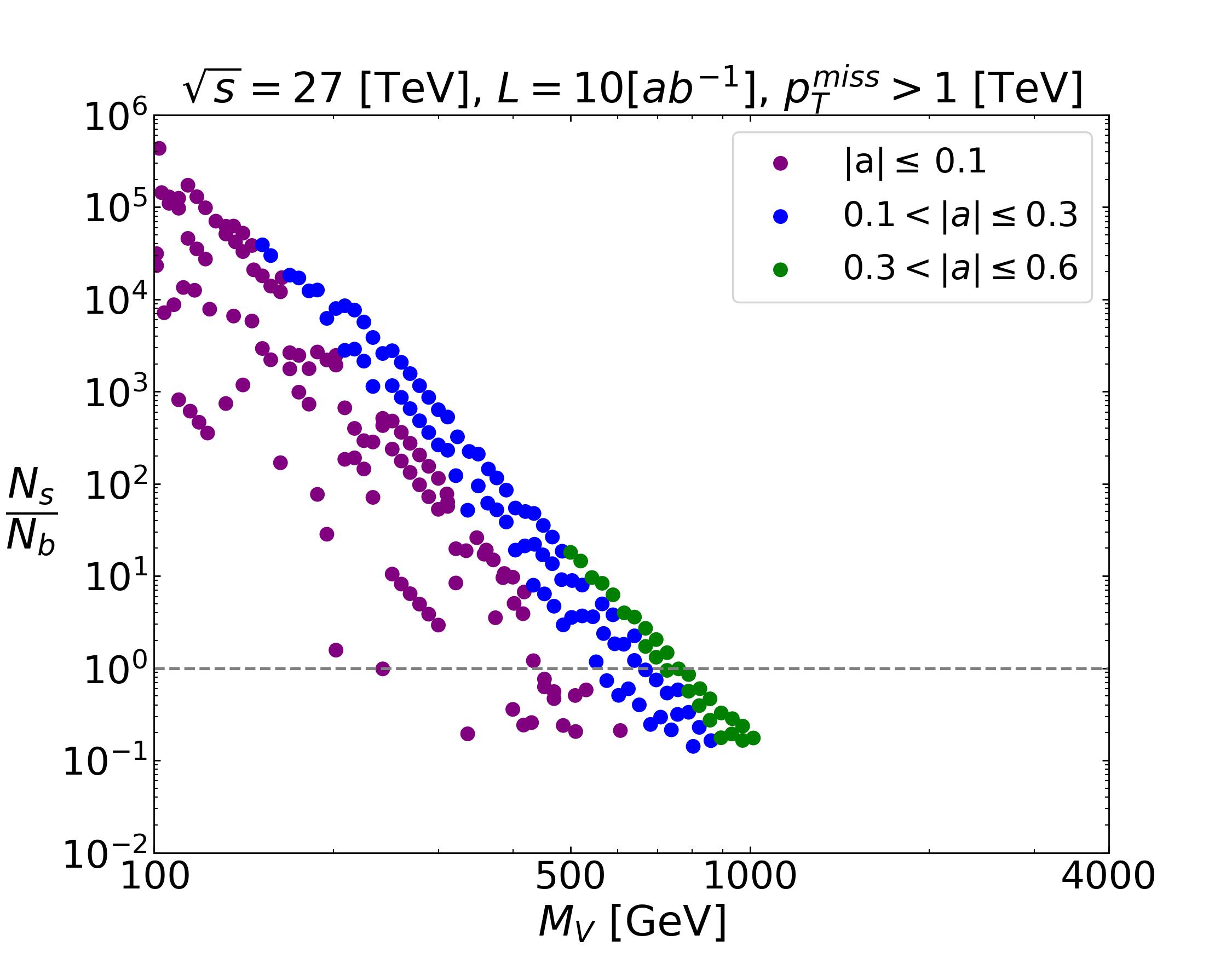}
    \includegraphics[width=0.373\linewidth]{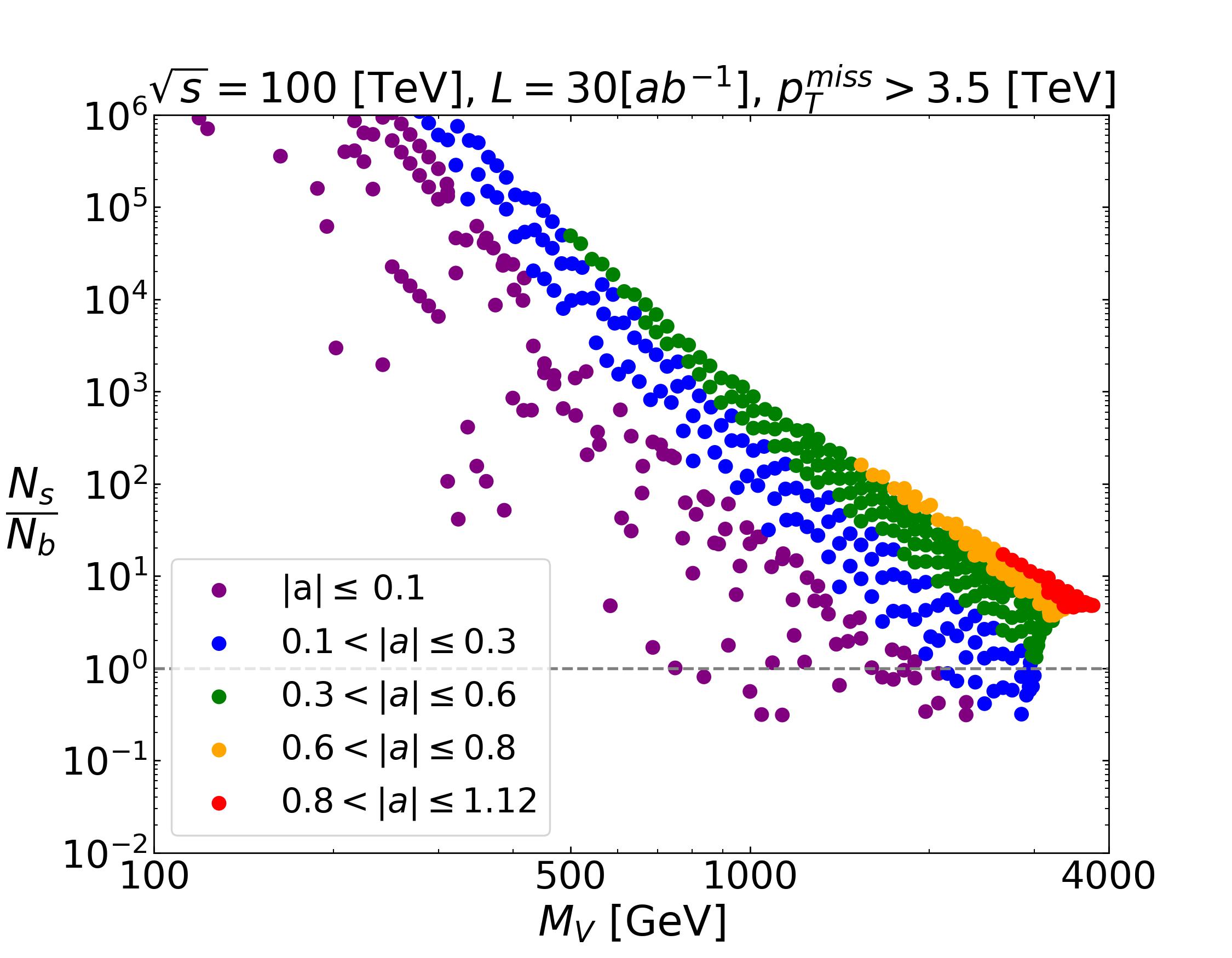}
    \caption{Ratio between signal and background events for each accelerator {\color{black}for the process} $pp\to hV^{+,0}V^{-,0}$. The {\color{black}dashed gray lines} indicates when the ratio is 1. All the points satisfy the constrain {\color{black}{$S>2$.}}}
    \label{events_excess_hVpVm}
\end{figure}

\begin{figure}[htbp]
    \centering
    \includegraphics[width=0.373\linewidth]{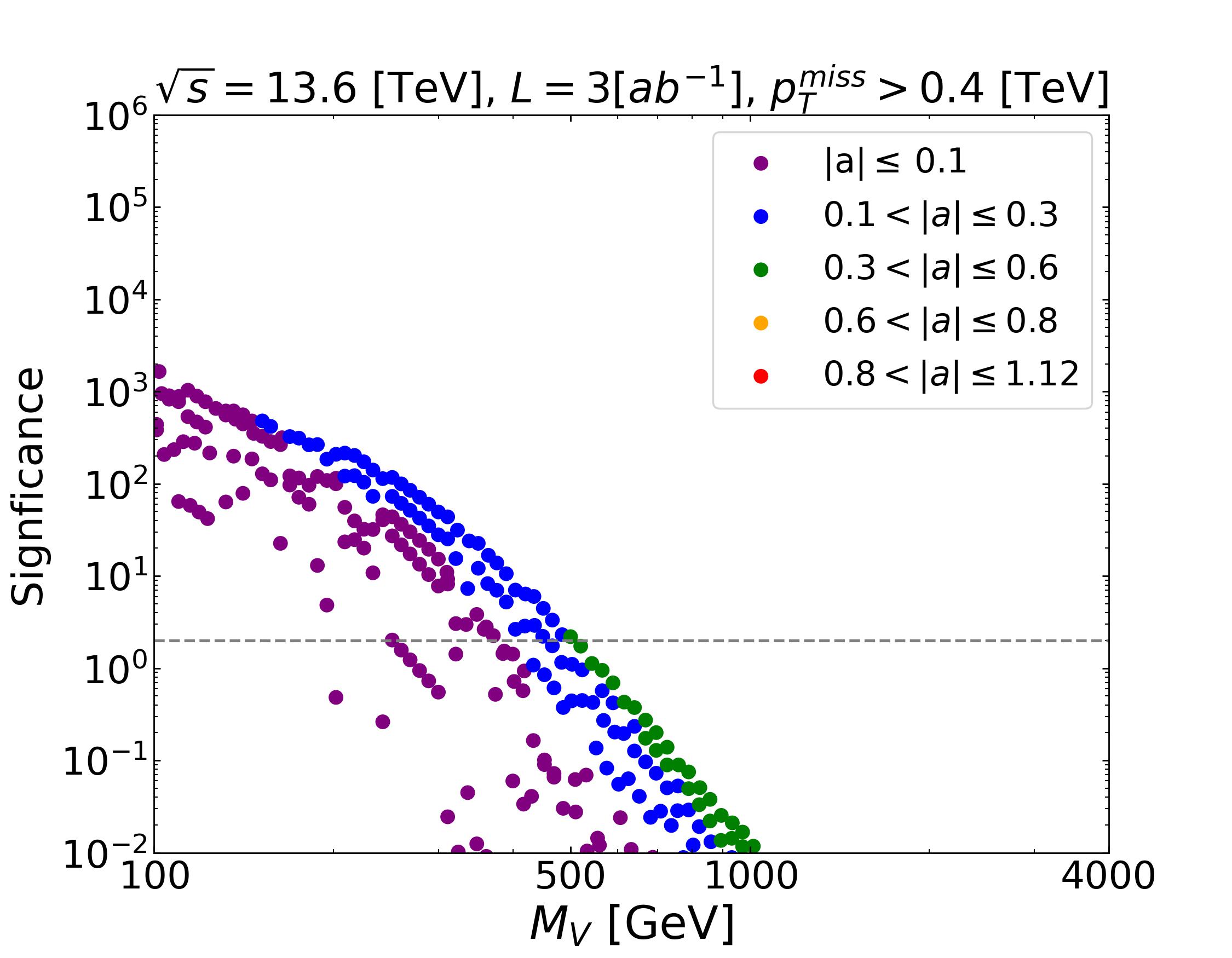}
    \includegraphics[width=0.373\linewidth]{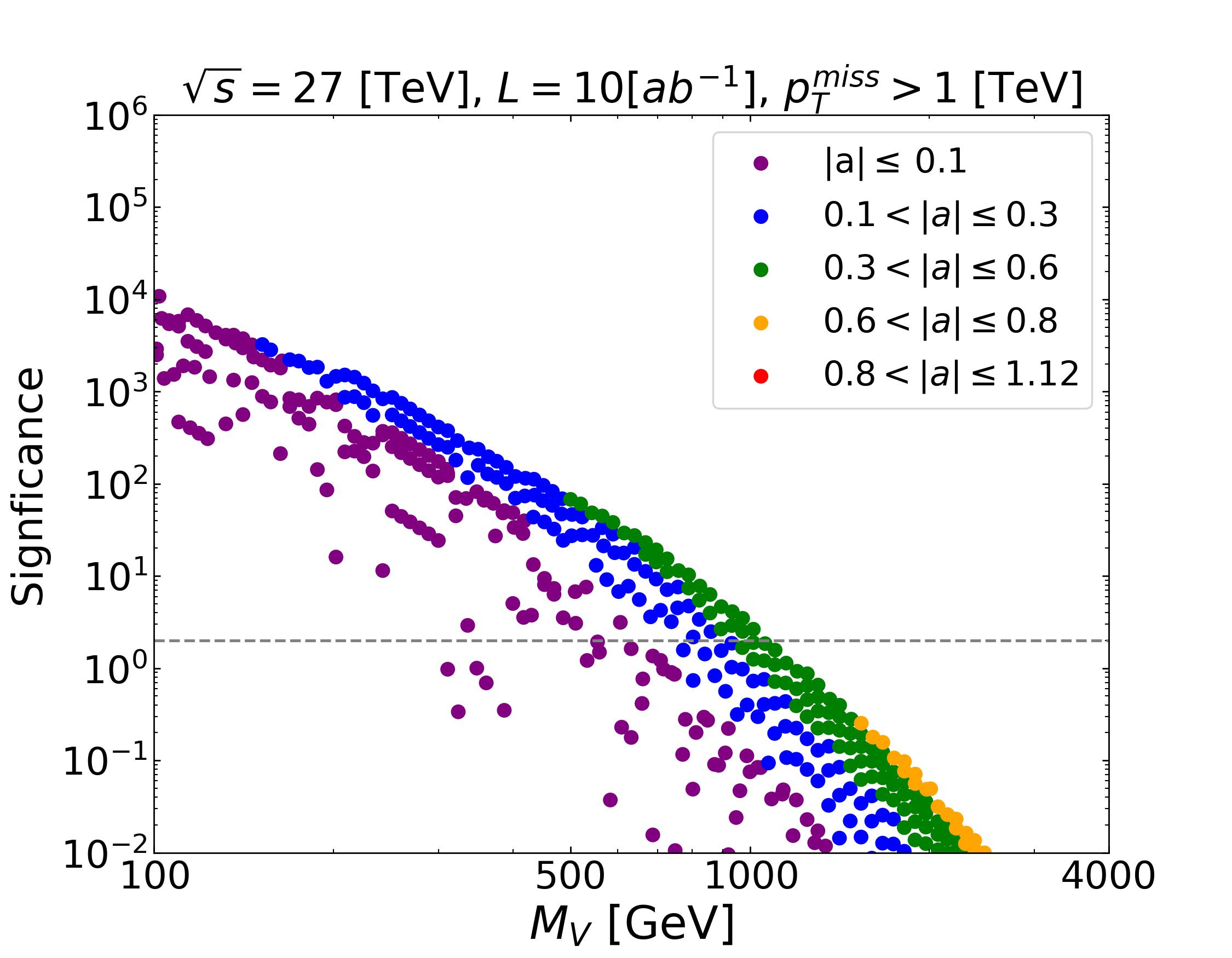}
    \includegraphics[width=0.373\linewidth]{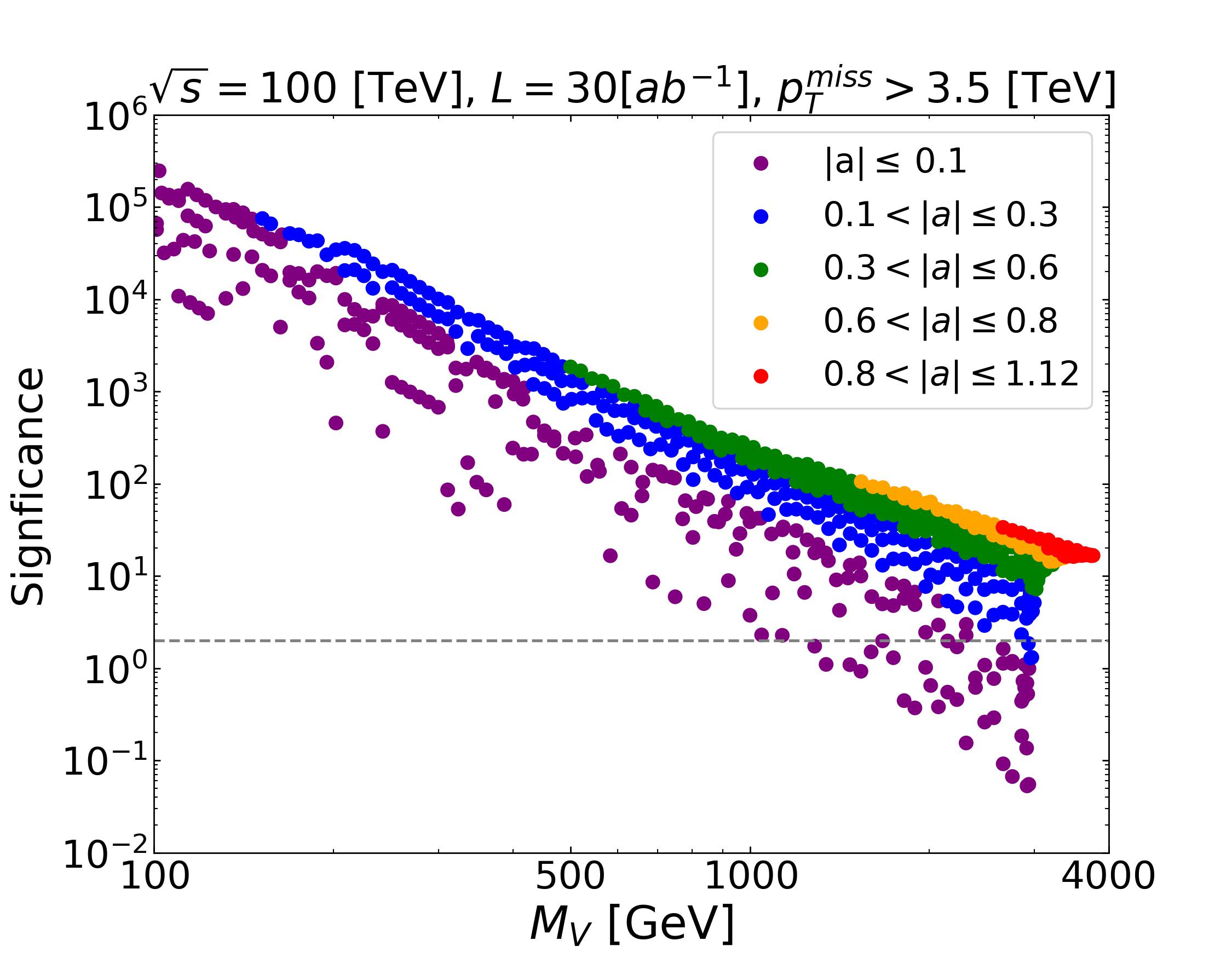}
    \caption{Statistical significance as function of the vector mass {\color{black}for the process} $pp\to hV^{+,0}V^{-,0}$. The {\color{black}dashed gray lines} indicates when the significance reach the 68\% confidence level ($S=2$).}
    \label{significance_hvpvm_sigmavsMv}
\end{figure}

\begin{figure}[htbp]
    \centering
    \includegraphics[width=0.373\linewidth]{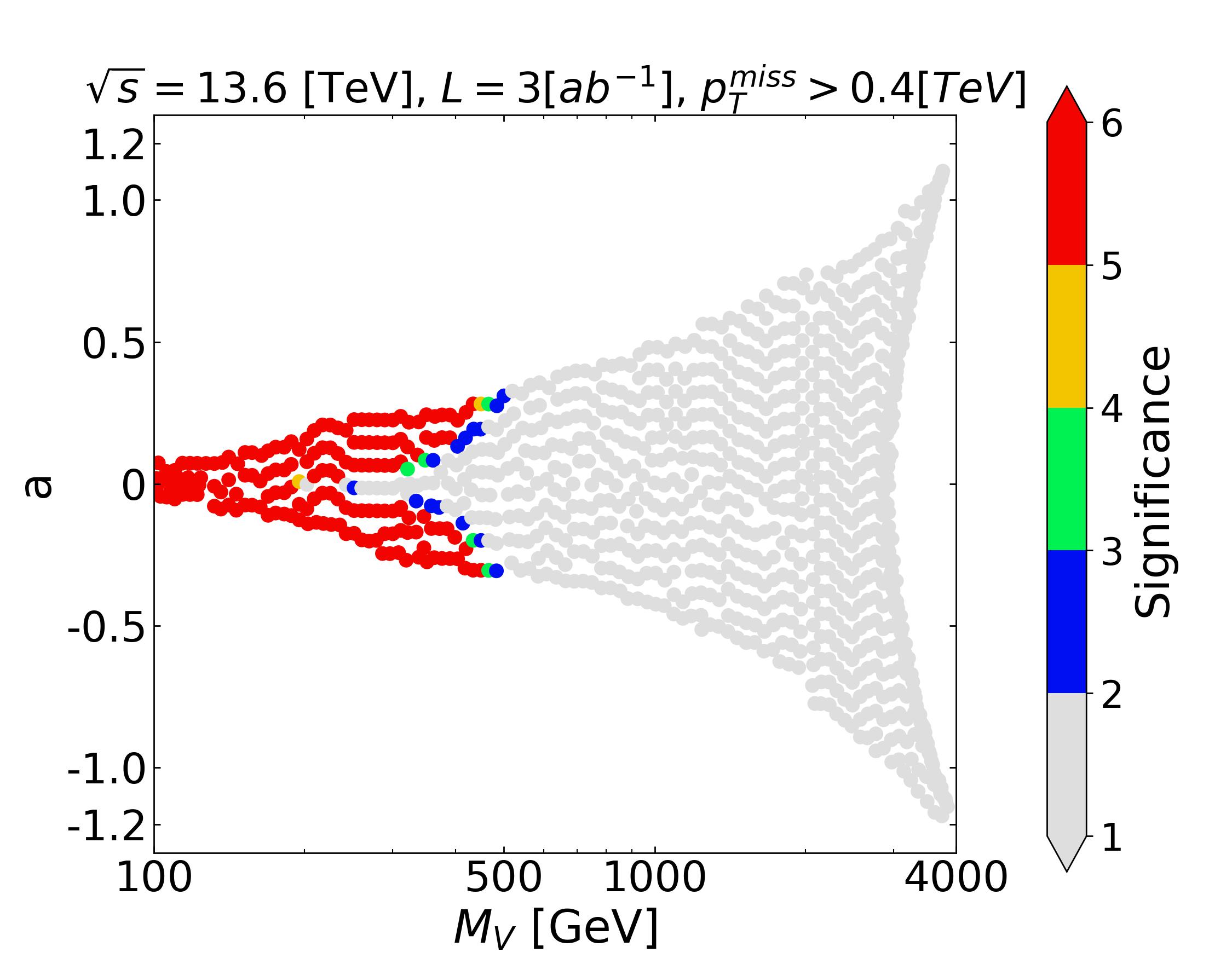}
    \includegraphics[width=0.373\linewidth]{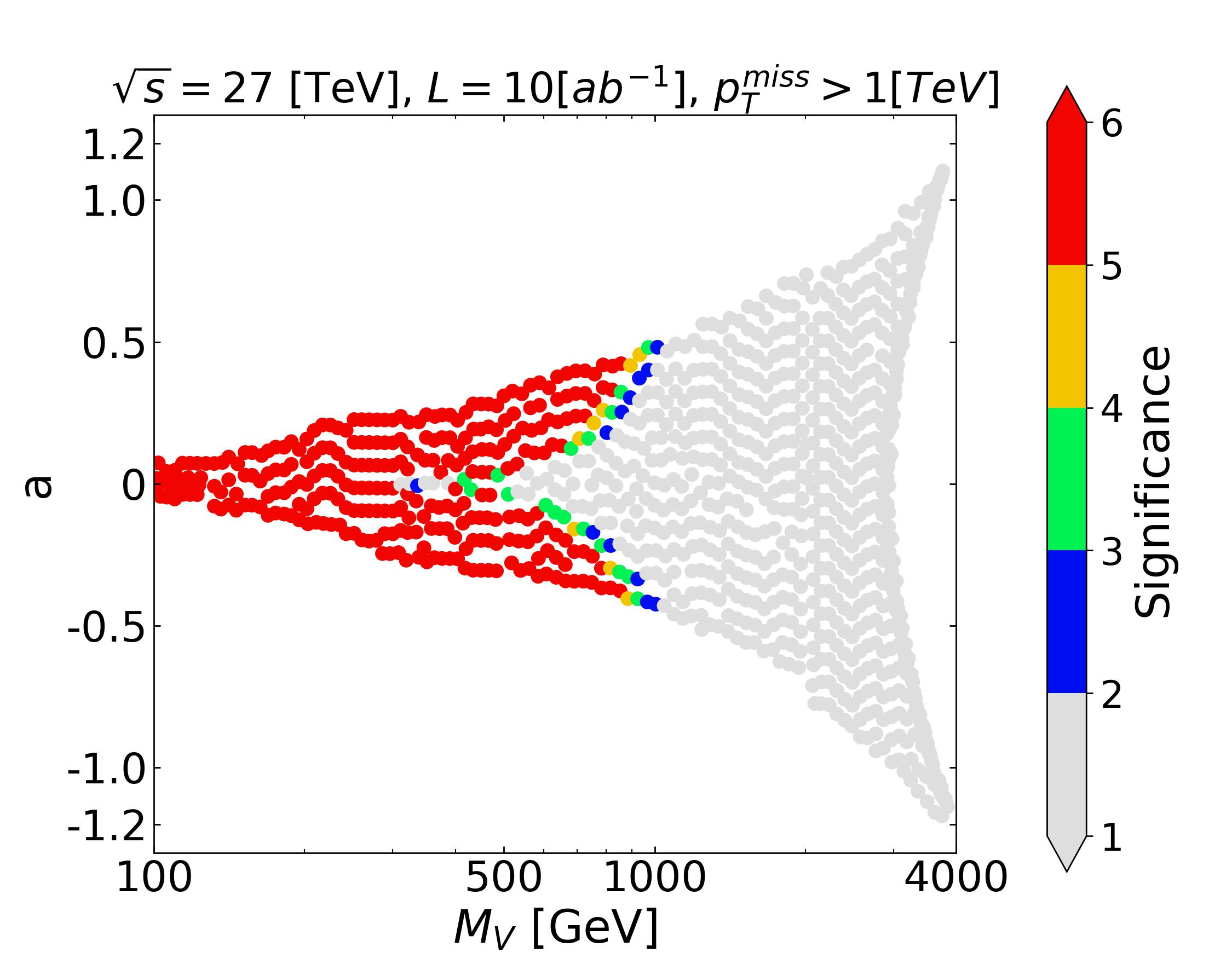}
    \includegraphics[width=0.373\linewidth]{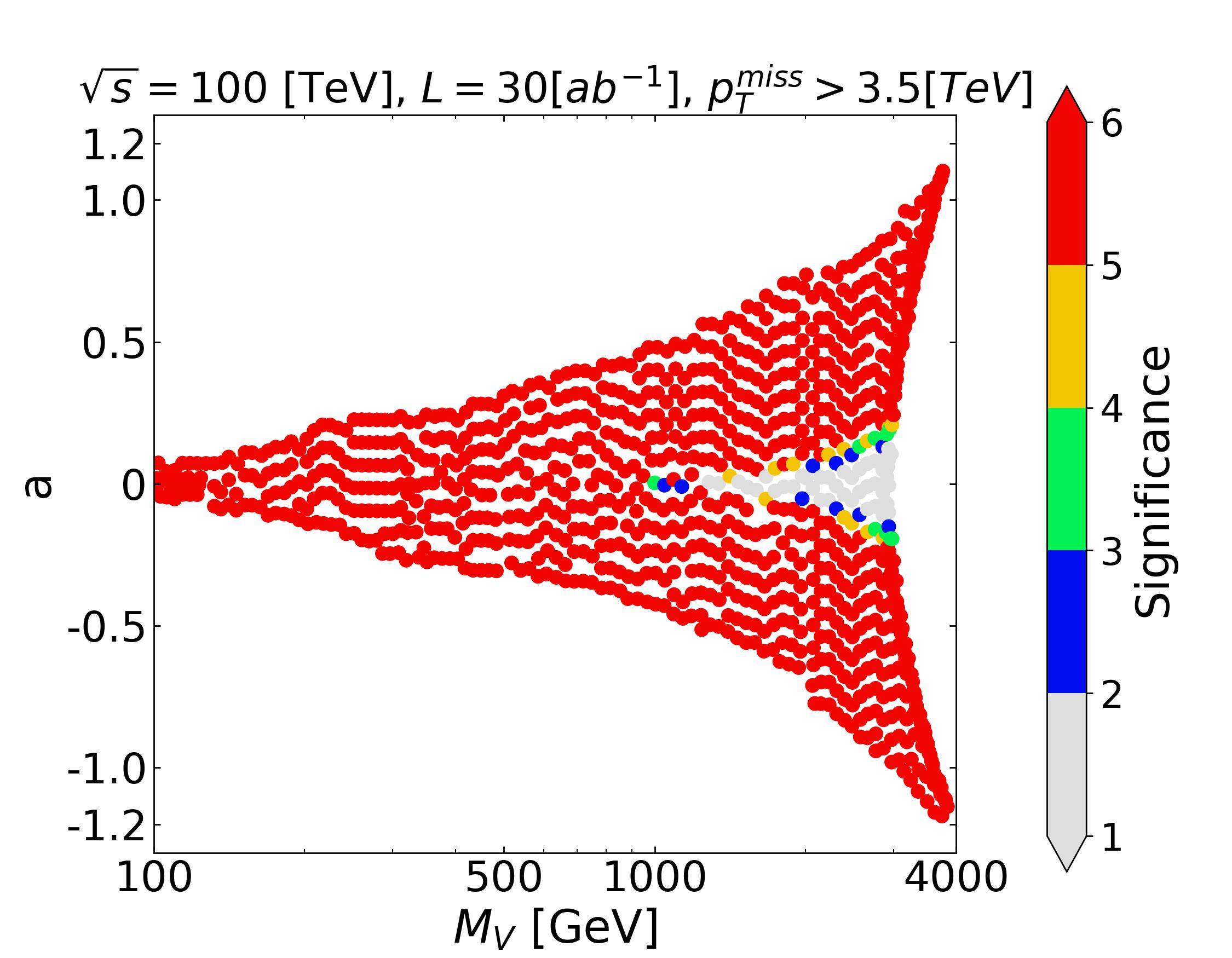}
    \caption{Parameter space showing the Significance in the color bar {\color{black}for the process} $pp\to hV^{+,0}V^{-,0}$ @ 13.6; 27 and 100 TeV of energy center of mass with its respective optimal kinematic cut and maximum integrated luminosity.}
    \label{significance_hvpvm_avsMv}
\end{figure}

However, as explained above, the charged companions for our DM candidate may also contribute to missing energy. In this case, the mono-Higgs production is originated from both: the $pp\to h V^0V^0$ and $pp\to h V^+V^-$ channels. The $pp\to h V^+V^-$ is favored by electroweak contributions that are absents $pp\to h V^0V^0$ , making the mono-Higgs production much stronger when the two sub-processes are taken into account. The difference {\color{black}is evident} when we compare Figure \ref{events_excess_hV0V0} and \ref{events_excess_hVpVm}: The signal to background ratio increase as much as four orders of magnitude at the low mass region and one order of magnitude for the highest masses. Of course, the statistical significance is also improved (see Figures \ref{significance_hvpvm_sigmavsMv} and \ref{significance_hvpvm_avsMv}).

\begin{figure}[htbp]
    \centering
    \includegraphics[width=0.373\linewidth]{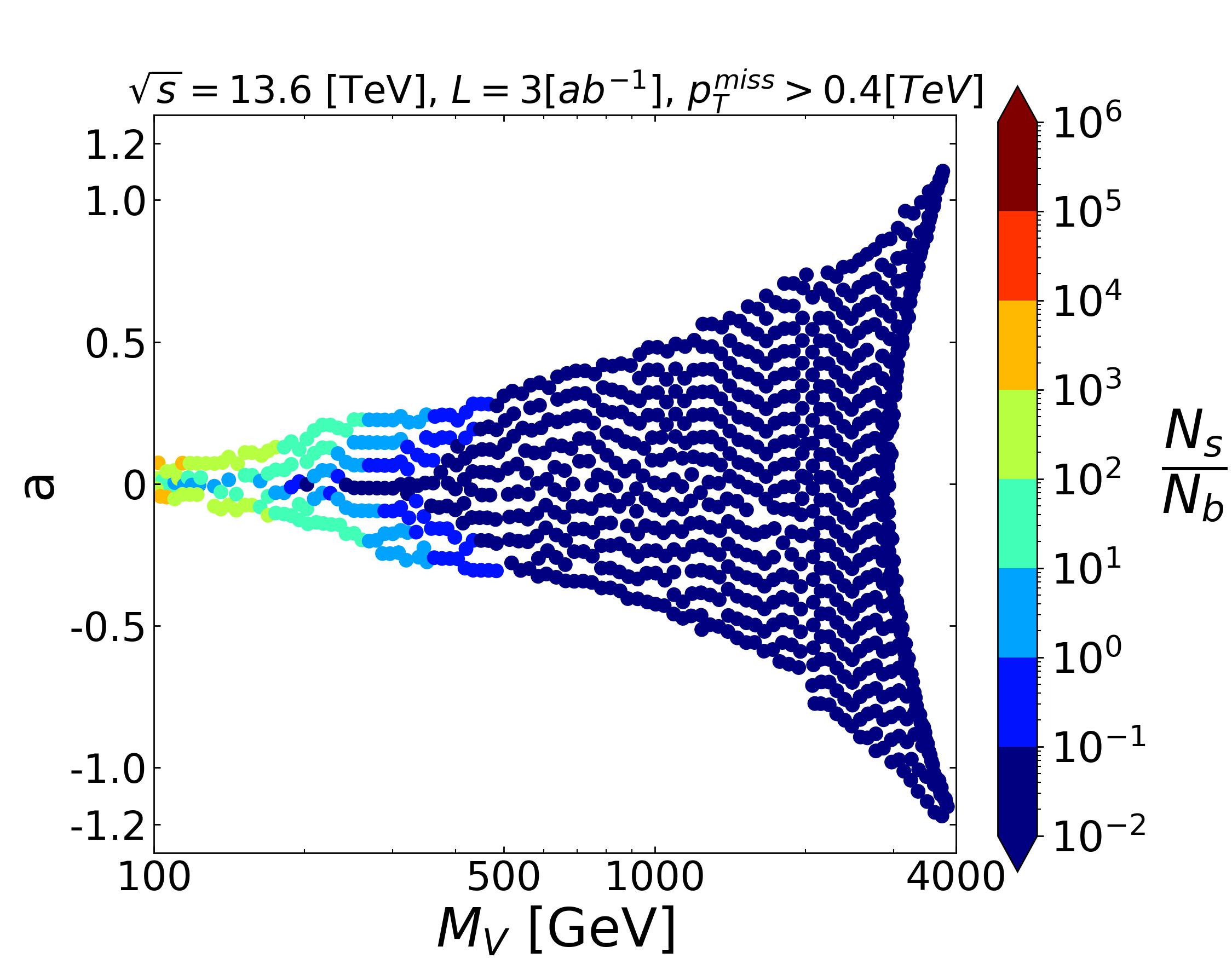}
    \includegraphics[width=0.373\linewidth]{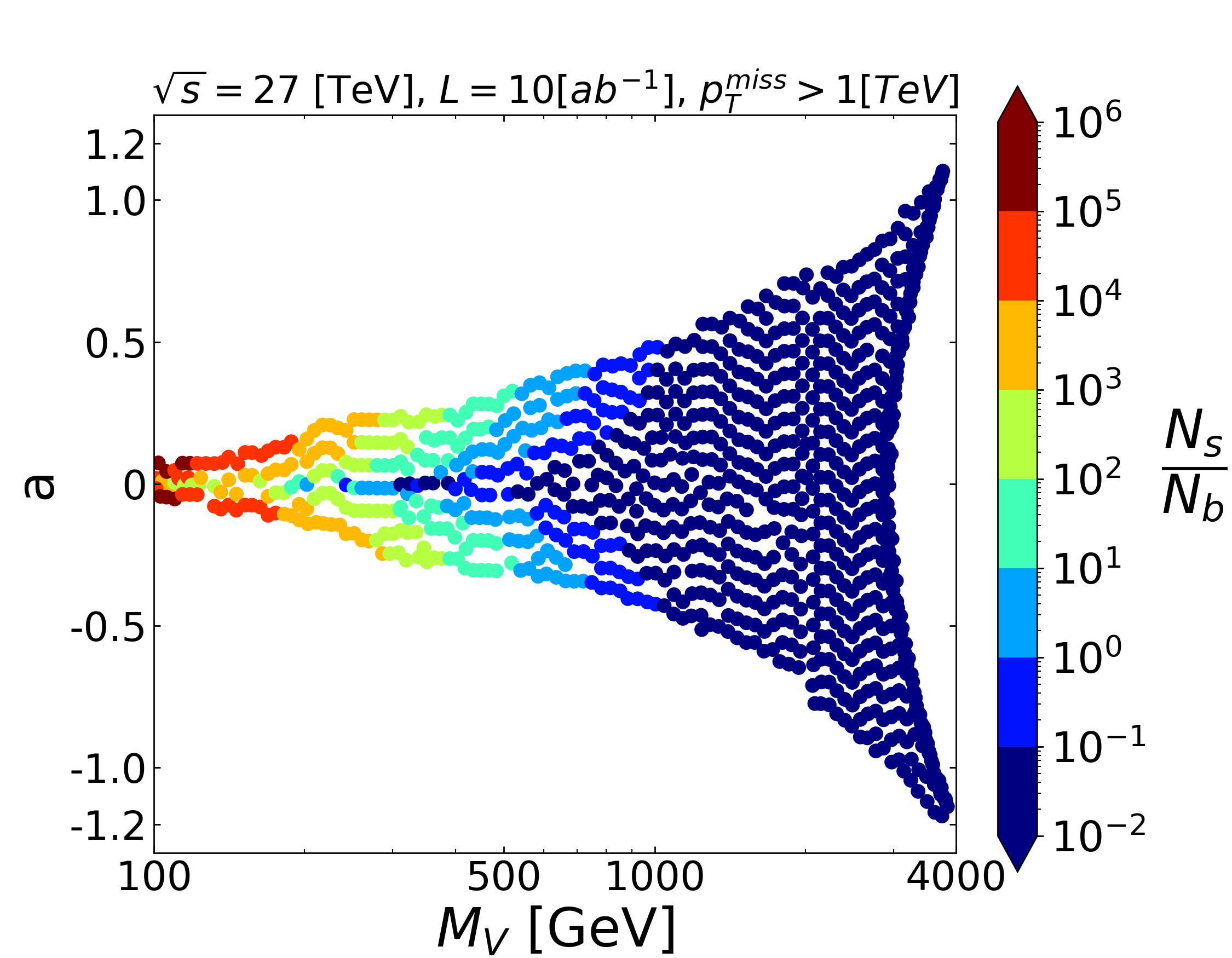}
    \includegraphics[width=0.373\linewidth]{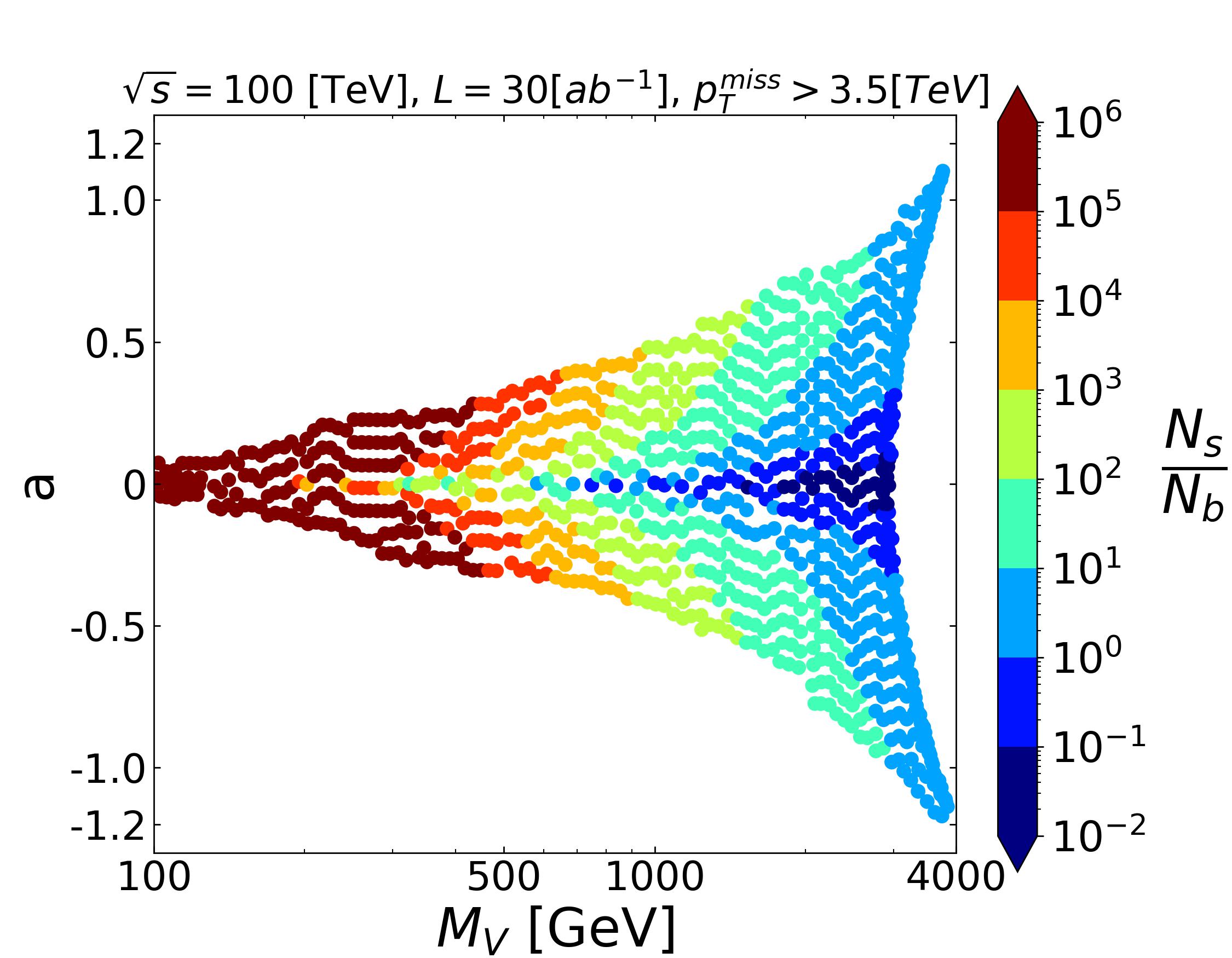}
    \caption{Parameter space displaying the value of the ratio between signal and background events in the color scale, for each accelerator {\color{black}for the process} $pp\to h V^{+,0} V^{-,0}$.}
    \label{events_excess_hV+V-_avsMv}
\end{figure}

Following the same conditions of Figures \ref{events_excess_hV0V0} and \ref{events_excess_hV0V0_avsMv}, in the Figures \ref{events_excess_hVpVm} and \ref{events_excess_hV+V-_avsMv} the excess of events in each collider is presented, assuming in each case the maximum projected luminosity. As the result of a greater signal-to-background ratio the HL-LHC can prove the model for $M_V \lesssim 500$ GeV and $|a|\lesssim$ {\color{black}$0.3$} (black and purple dots in Figure \ref{events_excess_hVpVm}). In the case of the HE-LHC, the model can be tested for masses as large as 1 TeV. 

The FCC-hh, on the other hand, can {\color{black}prove} most of parameter space, even part of the region  for which saturation of the relic density.

\clearpage
\newpage

\subsection{Mono-$Z$ Production}\label{subsec:Mono-Z_production}

Now, we turn our attention to the mono-$Z$ production process. Again, this process receives two contributions in our model: $pp\rightarrow V^0 V^0 Z$ and $pp\rightarrow V^{+,0} V^{-,0} Z$. Similarly to the analysis made for the mono-Higgs production, we study these two sub-processes separately.

\subsubsection{$V^0$ production}\label{subsubsec:Mono-Z_V0}

\begin{figure}[htbp]
    \centering
    \includegraphics[width=0.42\linewidth]{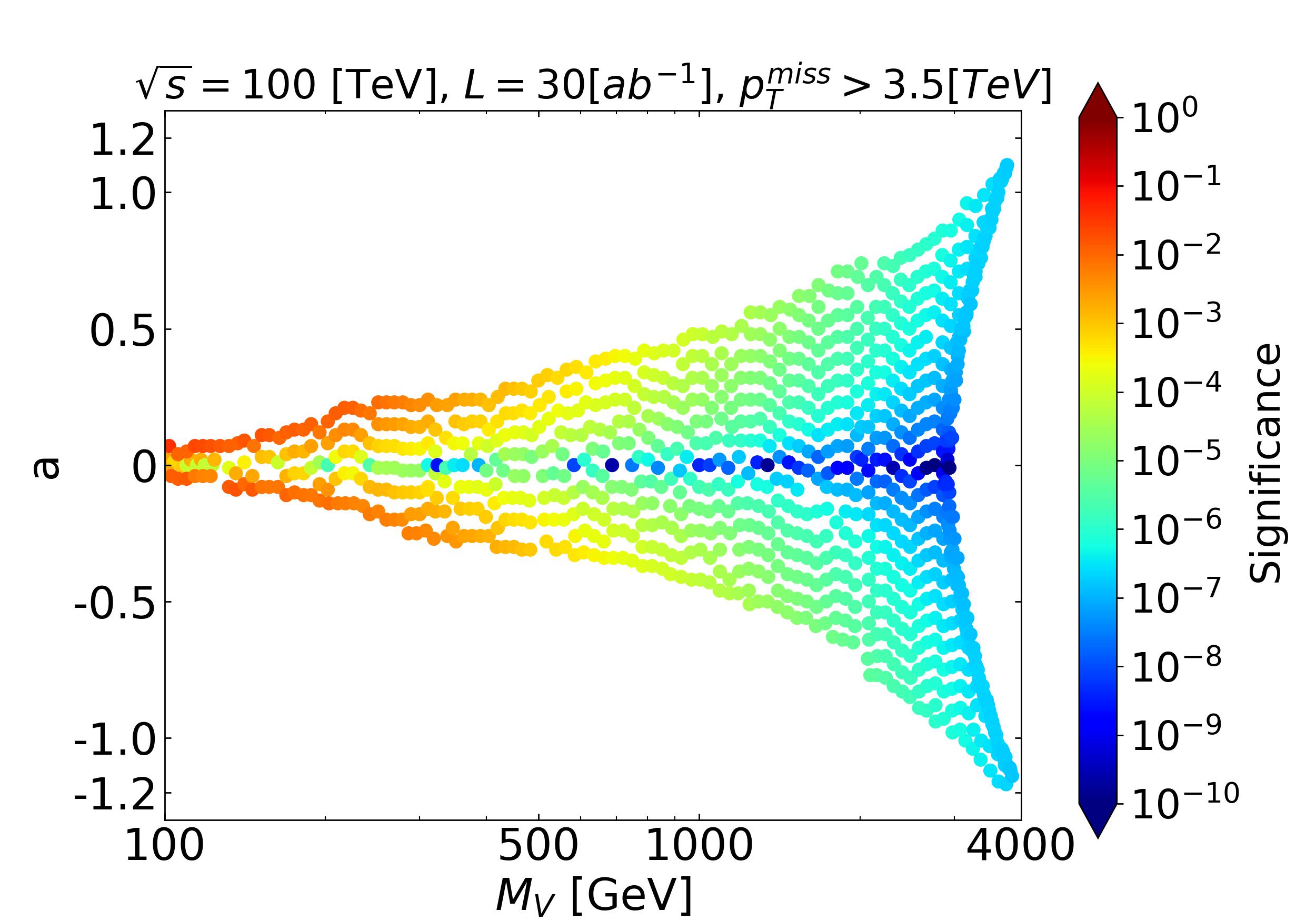}

    \caption{Statistical Significance in the parameter space for $pp\to ZV^0V^0$.}
    \label{fig:Mono-ZV0V0}
\end{figure}

In Figure \ref{fig:Mono-ZV0V0}, we show the behavior of statistical significance for the $pp\to ZV^0V^0$ at the FCC-hh with $L=30\,ab^{-1}$, which represents the highest expected luminosity for this collider. As we can see, the signal produced by this particular sub-process are not statistically significant even for the largest luminosity. Consequently, we will turn our attention toward the production of a $Z$ boson in association with the charged companions of the DM candidate.

\subsubsection{$V^{\pm,0}$ Production}\label{subsubsec:Mono-Z_Vpm}

\begin{figure}[htbp]
    \centering
    \includegraphics[width=0.373\linewidth]{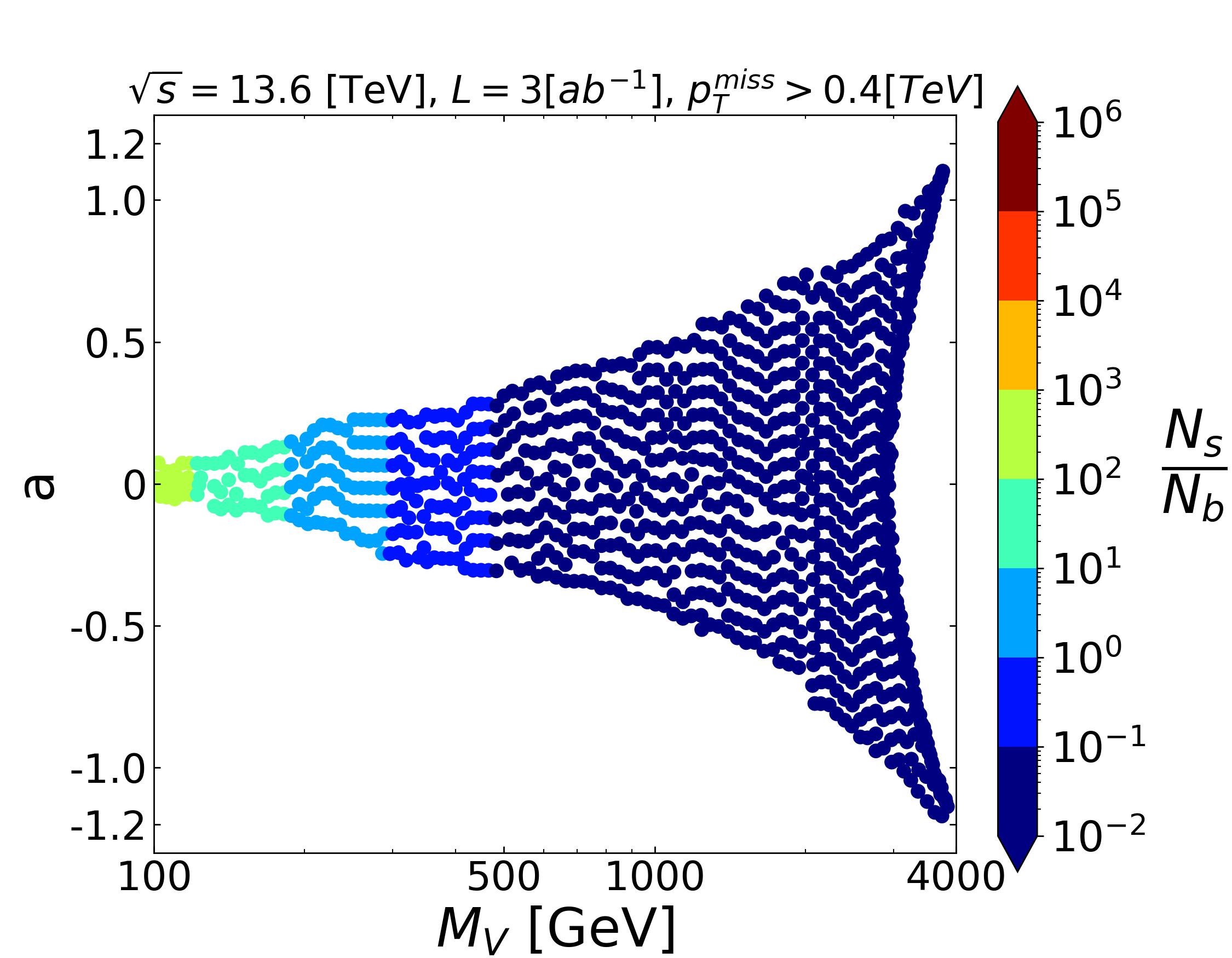}
    \includegraphics[width=0.373\linewidth]{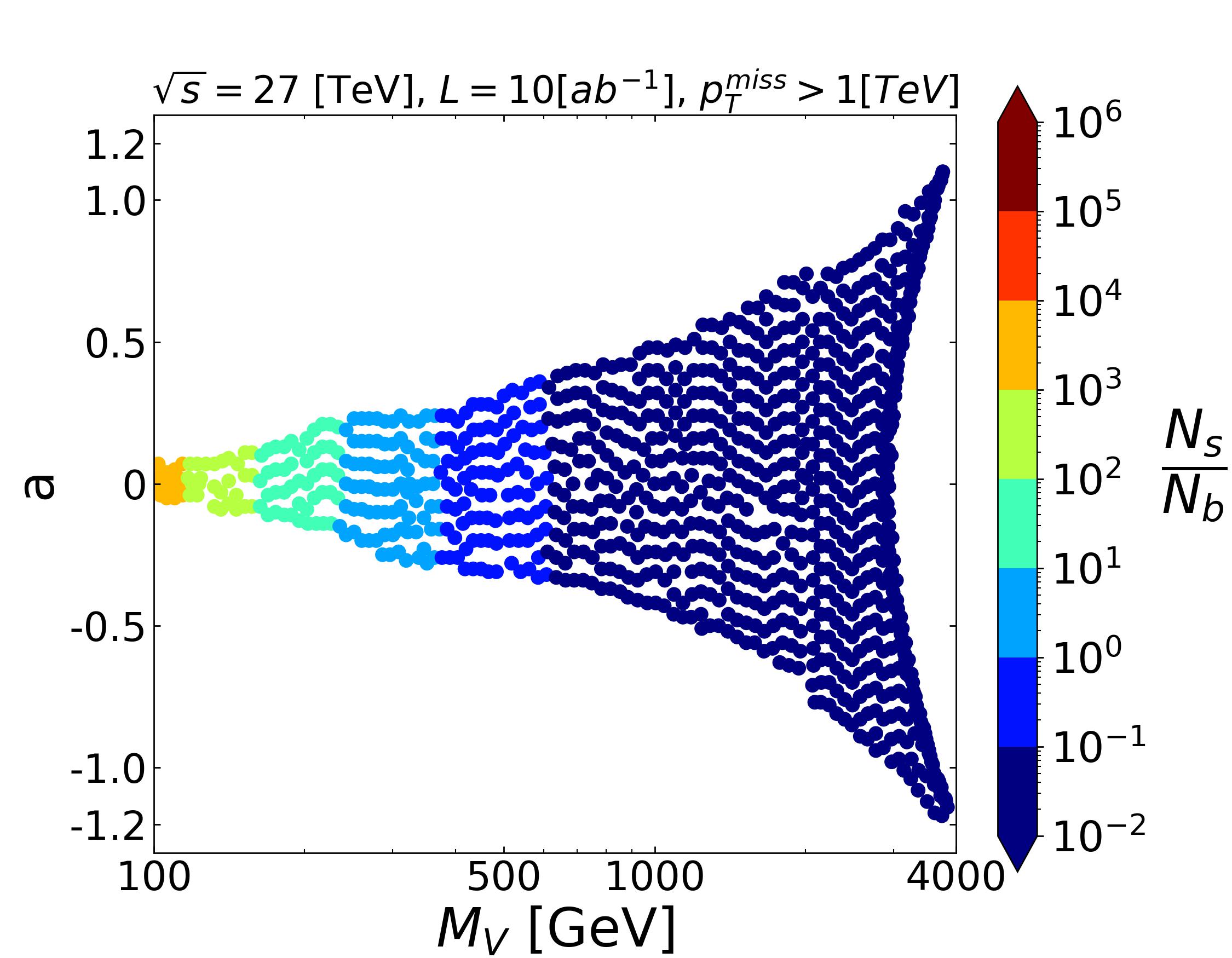}
    \includegraphics[width=0.373\linewidth]{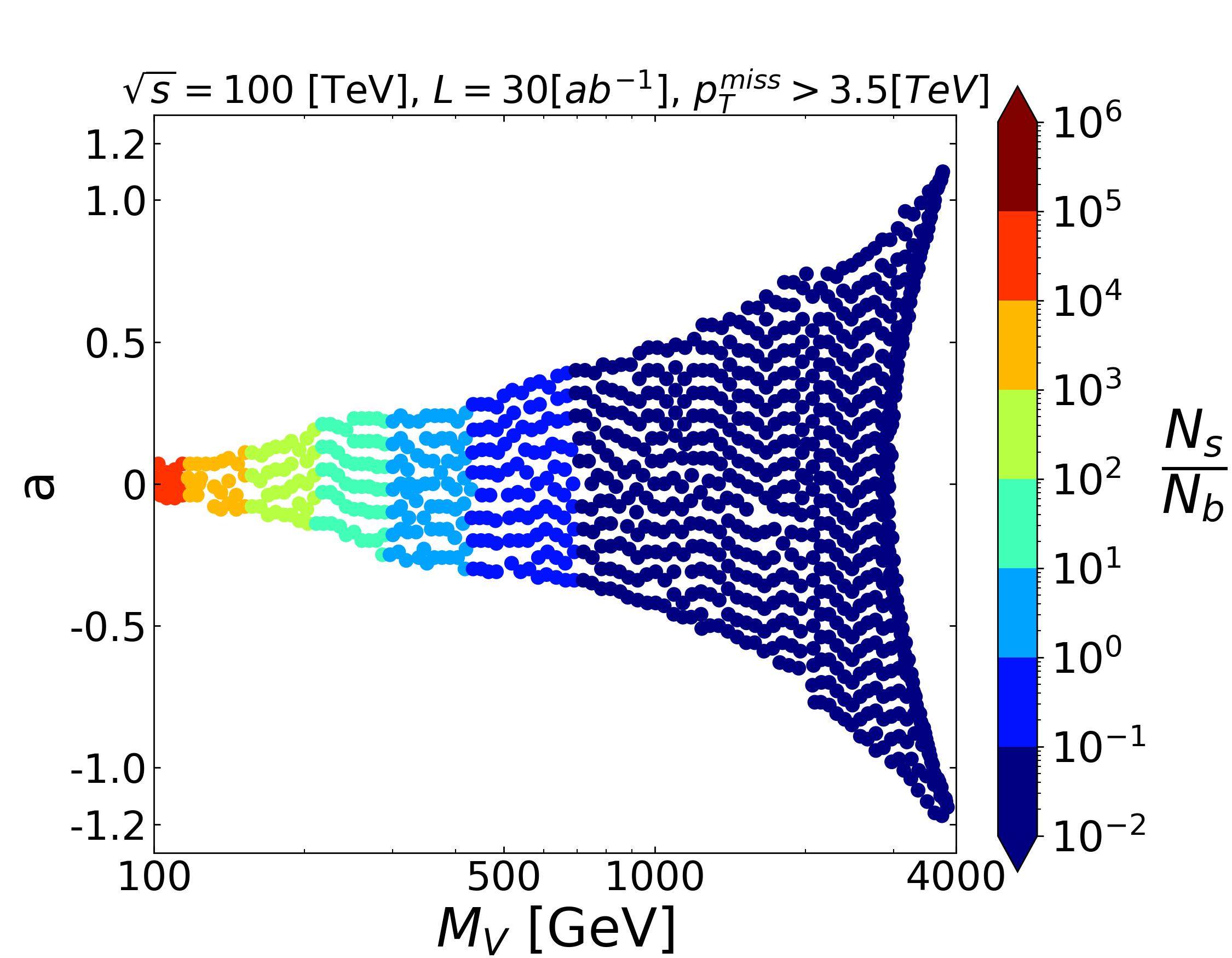}
    \caption{Parameter space displaying the value of the ratio between signal and background events in the color scale, for each accelerator {\color{black}for the process} $pp\to ZV^{+,0} V^{-,0}$.}
    \label{events_excess_ZV+V-_avsMv}
\end{figure}

\begin{figure}[htbp]
    \centering
    \includegraphics[width=0.373\linewidth]{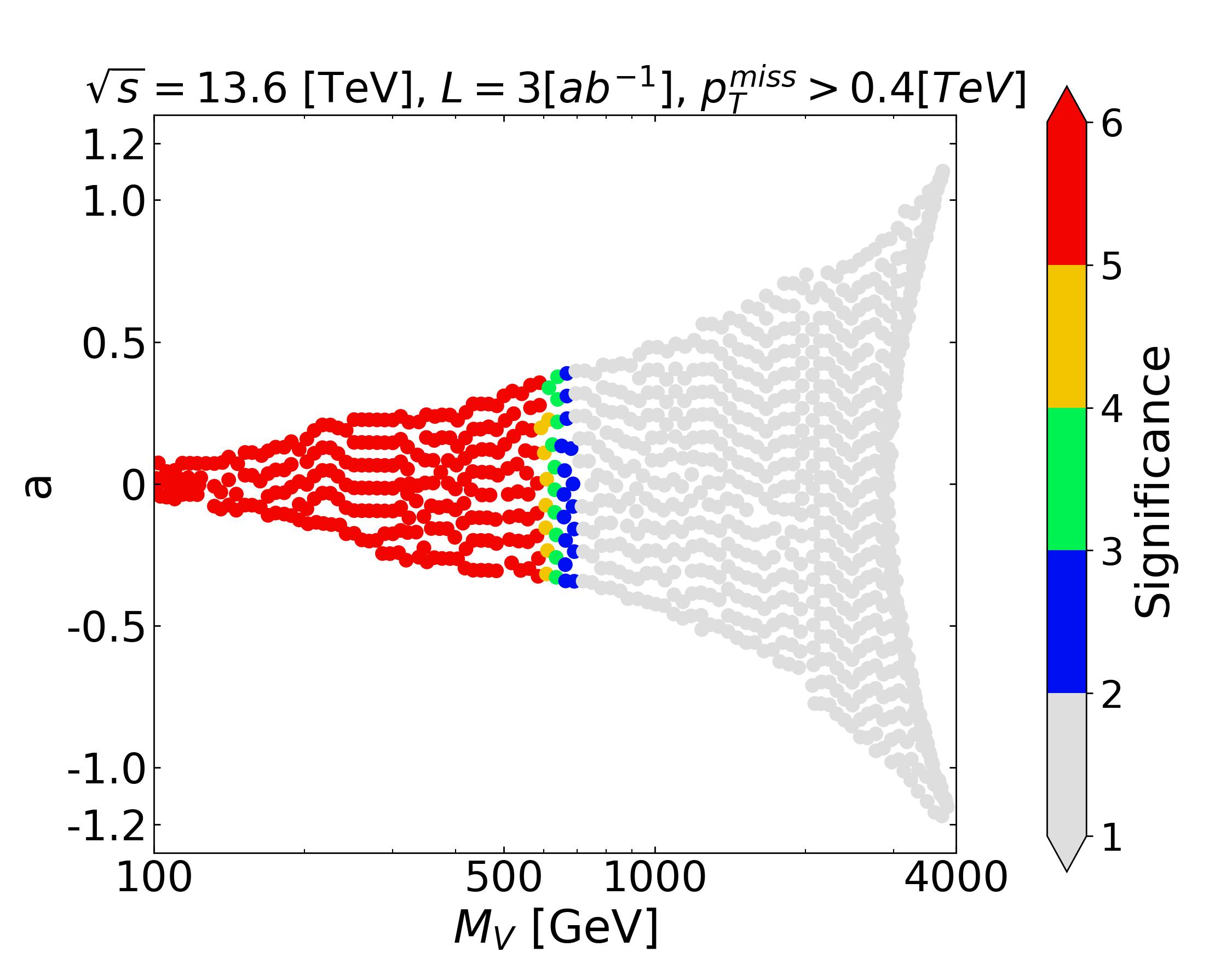}
    \includegraphics[width=0.373\linewidth]{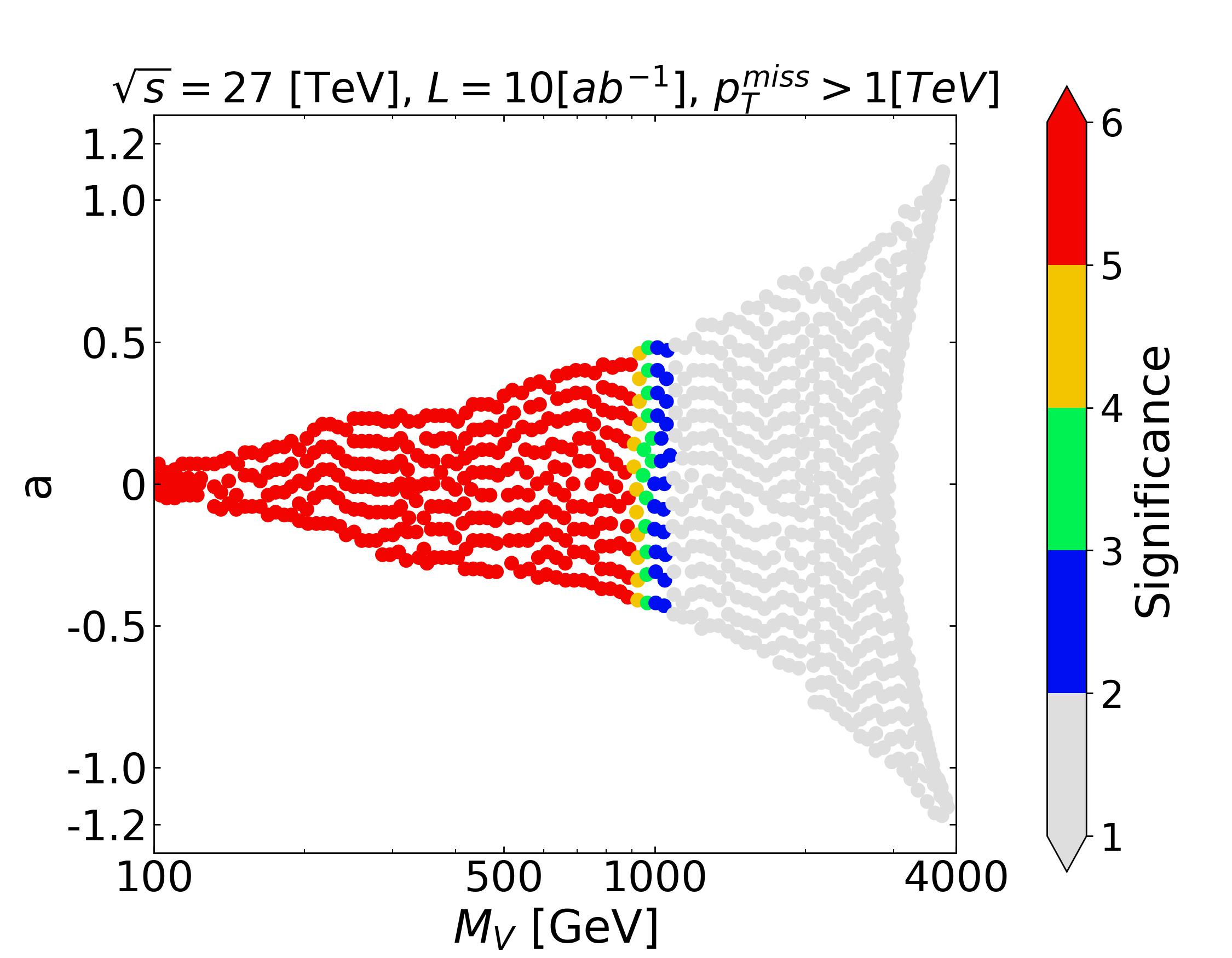}
    \includegraphics[width=0.373\linewidth]{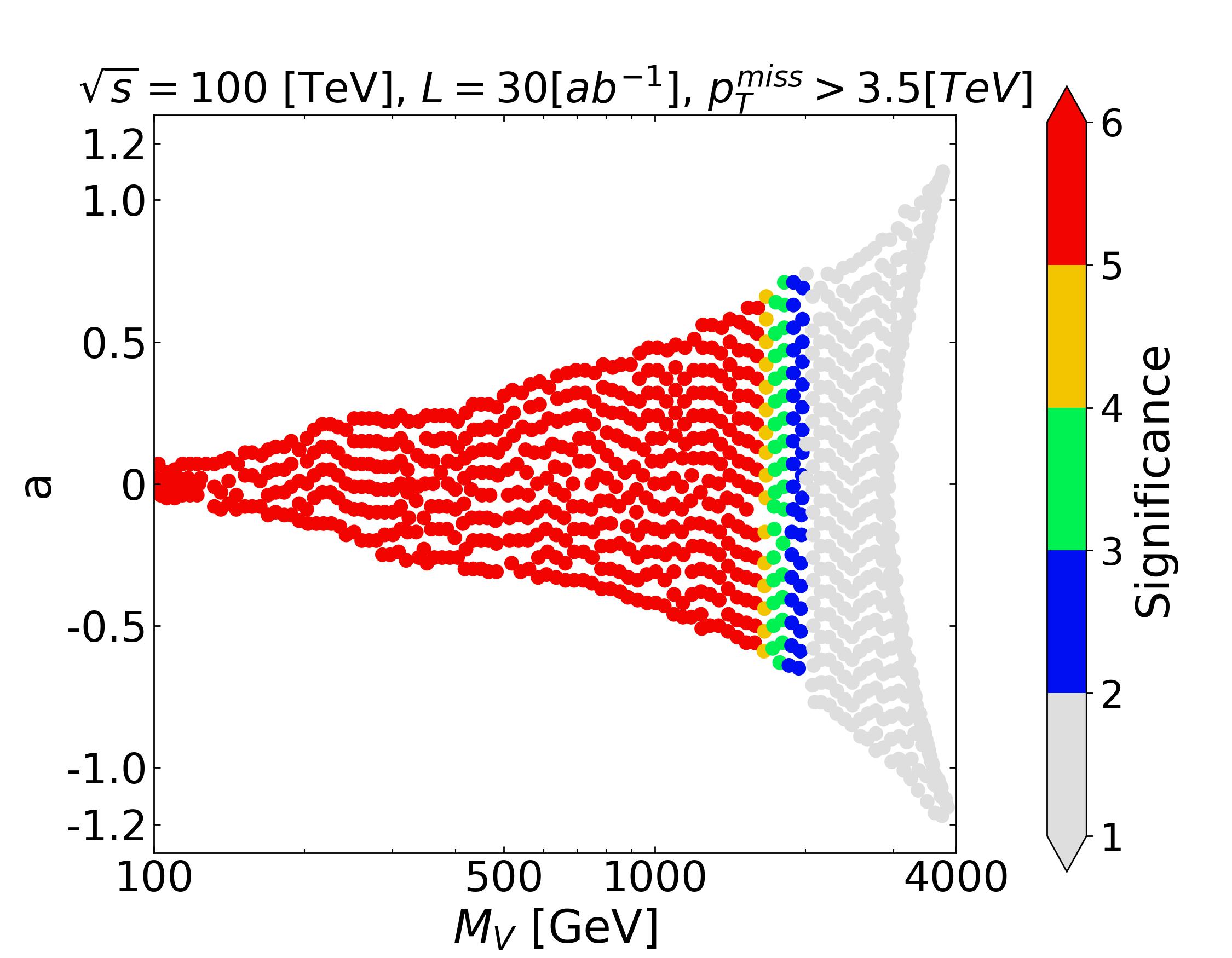}
    \caption{Parameter space showing the Significance in the color bar {\color{black}for the process} $pp\to ZV^{+,0}V^{-,0}$ @ 13.6; 27 and 100 TeV of energy center of mass with its respective optimal kinematic cut and maximum integrated luminosity.}
    \label{significance_Zvpvm_avsMv}
\end{figure}

Although we are faced with the same large background as before, here the signal is much powerful due to the electroweak channels. As we did in the mono-Higgs production case, we study the signal-to-background ratio (Figure \ref{events_excess_ZV+V-_avsMv}) and its statistical significance (Figure \ref{significance_Zvpvm_avsMv}) at different colliders. This time, significant excesses of events are produced in most parts of the allowed parameter space, making the model testable through this channel  for masses up to a few TeV.

The main feature of the Mono-$Z$ event is its independence of the Higgs-DM coupling parameter $(a)$, {\color{black}thus} the mass of the vector $(M_V)$ {\color{black}is} the only important parameter to consider. In this sense, the mono-$Z$ production is complementary to the mono-Higgs channel.

\clearpage
\newpage

\section{Conclusions}\label{Sec:Conclusions}
In this work, we have studied the mono-Higgs and the mono-$Z$ production channels in the Minimal Vector Dark Matter model. We argue that the production of the Higgs and the $Z$ boson in association with the charged companions of the DM candidate must be taken into account and contribute significantly to the studied channels. The study of the mono-Higgs production at the FCC may, in principle, test the whole parameter space with the exception of a small region at high DM masses and very low Higgs-DM coupling. For this reason, the mono-Higgs channel may be represent the best opportunity to discover or rule out this model at future colliders.  Although the mono-Higgs  is always the most important channel, the mono-$Z$ is complementary and in the case of discovery in the mono-Higgs, it may help in determining the values of the parameters of the model. 

\

\begin{acknowledgments*}
This work was partially financed by Fondecyt grant 1230110, ANID PIA/APOYO AFB220004, financial support from UTFSM master scholarship No. 048/2022 and ANID—The Millennium Science Initiative Program ICN2019\_044.
\end{acknowledgments*}

\section*{References}
\bibliographystyle{unsrtnat}
\bibliography{references}

\begin{thebibliography}{27}
\providecommand{\natexlab}[1]{#1}
\providecommand{\url}[1]{\texttt{#1}}
\expandafter\ifx\csname urlstyle\endcsname\relax
  \providecommand{\doi}[1]{doi: #1}\else
  \providecommand{\doi}{doi: \begingroup \urlstyle{rm}\Url}\fi

\bibitem[Cirelli et~al.(2006)Cirelli, Fornengo, and Strumia]{Cirelli2006}
Marco Cirelli, Nicolao Fornengo, and Alessandro Strumia.
\newblock {Minimal dark matter}.
\newblock \emph{Nucl. Phys. B}, 753:\penalty0 178--194, 2006.
\newblock \doi{10.1016/j.nuclphysb.2006.07.012}.

\bibitem[S\'aez et~al.(2019)S\'aez, Rojas-Abatte, and Zerwekh]{Saez:fundamental}
Bastian~D\'\i{}az S\'aez, Felipe Rojas-Abatte, and Alfonso~R. Zerwekh.
\newblock {Dark Matter from a Vector Field in the Fundamental Representation of $SU(2)_L$}.
\newblock \emph{Phys. Rev. D}, 99\penalty0 (7):\penalty0 075026, 2019.
\newblock \doi{10.1103/PhysRevD.99.075026}.

\bibitem[Belyaev et~al.(2019)Belyaev, Cacciapaglia, Mckay, Marin, and Zerwekh]{vtdm:adjoint}
Alexander Belyaev, Giacomo Cacciapaglia, James Mckay, Dixon Marin, and Alfonso~R. Zerwekh.
\newblock {Minimal Spin-one Isotriplet Dark Matter}.
\newblock \emph{Phys. Rev. D}, 99\penalty0 (11):\penalty0 115003, 2019.
\newblock \doi{10.1103/PhysRevD.99.115003}.

\bibitem[Abe et~al.(2020)Abe, Fujiwara, Hisano, and Matsushita]{Abe:2020mph}
Tomohiro Abe, Motoko Fujiwara, Junji Hisano, and Kohei Matsushita.
\newblock {A model of electroweakly interacting non-abelian vector dark matter}.
\newblock \emph{JHEP}, 07:\penalty0 136, 2020.
\newblock \doi{10.1007/JHEP07(2020)136}.

\bibitem[Zerwekh(2013)]{Zerwekh:2012bf}
Alfonso~R. Zerwekh.
\newblock {On the Quantum Chromodynamics of a Massive Vector Field in the Adjoint Representation}.
\newblock \emph{Int. J. Mod. Phys. A}, 28:\penalty0 1350054, 2013.
\newblock \doi{10.1142/S0217751X13500541}.

\bibitem[Aprile et~al.(2020)]{XENON:2020kmp}
E.~Aprile et~al.
\newblock {Projected WIMP sensitivity of the XENONnT dark matter experiment}.
\newblock \emph{JCAP}, 11:\penalty0 031, 2020.
\newblock \doi{10.1088/1475-7516/2020/11/031}.

\bibitem[Belyaev et~al.(2021)Belyaev, Prestel, Rojas-Abbate, and Zurita]{Belyaev:2020dissapearing}
Alexander Belyaev, Stefan Prestel, Felipe Rojas-Abbate, and Jose Zurita.
\newblock {Probing dark matter with disappearing tracks at the LHC}.
\newblock \emph{Phys. Rev. D}, 103\penalty0 (9):\penalty0 095006, 2021.
\newblock \doi{10.1103/PhysRevD.103.095006}.

\bibitem[Aghanim et~al.(2020)]{Planck:2018vyg}
N.~Aghanim et~al.
\newblock {Planck 2018 results. VI. Cosmological parameters}.
\newblock \emph{Astron. Astrophys.}, 641:\penalty0 A6, 2020.
\newblock \doi{10.1051/0004-6361/201833910}.
\newblock [Erratum: Astron.Astrophys. 652, C4 (2021)].

\bibitem[Fukuda et~al.(2018)Fukuda, Nagata, Otono, and Shirai]{Fukuda:2017disstrackcharged}
Hajime Fukuda, Natsumi Nagata, Hidetoshi Otono, and Satoshi Shirai.
\newblock {Higgsino Dark Matter or Not: Role of Disappearing Track Searches at the LHC and Future Colliders}.
\newblock \emph{Phys. Lett. B}, 781:\penalty0 306--311, 2018.
\newblock \doi{10.1016/j.physletb.2018.03.088}.

\bibitem[Saito et~al.(2019)Saito, Sawada, Terashi, and Asai]{Saito:2019disstrackcharged}
Masahiko Saito, Ryu Sawada, Koji Terashi, and Shoji Asai.
\newblock {Discovery reach for wino and higgsino dark matter with a disappearing track signature at a 100 TeV $pp$ collider}.
\newblock \emph{Eur. Phys. J. C}, 79\penalty0 (6):\penalty0 469, 2019.
\newblock \doi{10.1140/epjc/s10052-019-6974-2}.

\bibitem[Mahbubani et~al.(2017)Mahbubani, Schwaller, and Zurita]{Mahbubani:2017disstrackcharged}
Rakhi Mahbubani, Pedro Schwaller, and Jose Zurita.
\newblock {Closing the window for compressed Dark Sectors with disappearing charged tracks}.
\newblock \emph{JHEP}, 06:\penalty0 119, 2017.
\newblock \doi{10.1007/JHEP06(2017)119}.
\newblock [Erratum: JHEP 10, 061 (2017)].

\bibitem[Ghorbani and Khalkhali(2017)]{Ghorbani:2016edw}
Karim Ghorbani and Leila Khalkhali.
\newblock {Mono-Higgs signature in a fermionic dark matter model}.
\newblock \emph{J. Phys. G}, 44\penalty0 (10):\penalty0 105004, 2017.
\newblock \doi{10.1088/1361-6471/aa823a}.

\bibitem[Aad et~al.(2021)]{ATLAS:2021shl}
Georges Aad et~al.
\newblock {Search for dark matter produced in association with a Standard Model Higgs boson decaying into b-quarks using the full Run 2 dataset from the ATLAS detector}.
\newblock \emph{JHEP}, 11:\penalty0 209, 2021.
\newblock \doi{10.1007/JHEP11(2021)209}.

\bibitem[No(2016)]{No:2015xqa}
Jose~Miguel No.
\newblock {Looking through the pseudoscalar portal into dark matter: Novel mono-Higgs and mono-Z signatures at the LHC}.
\newblock \emph{Phys. Rev. D}, 93\penalty0 (3):\penalty0 031701, 2016.
\newblock \doi{10.1103/PhysRevD.93.031701}.

\bibitem[Carpenter et~al.(2014)Carpenter, DiFranzo, Mulhearn, Shimmin, Tulin, and Whiteson]{Carpenter:2013xra}
Linda Carpenter, Anthony DiFranzo, Michael Mulhearn, Chase Shimmin, Sean Tulin, and Daniel Whiteson.
\newblock {Mono-Higgs-boson: A new collider probe of dark matter}.
\newblock \emph{Phys. Rev. D}, 89\penalty0 (7):\penalty0 075017, 2014.
\newblock \doi{10.1103/PhysRevD.89.075017}.

\bibitem[Bhowmik et~al.(2022)Bhowmik, Lahiri, Bhattacharya, Mukhopadhyaya, and Singh]{Bhowmik:2020spw}
Debabrata Bhowmik, Jayita Lahiri, Satyaki Bhattacharya, Biswarup Mukhopadhyaya, and Ritesh~K. Singh.
\newblock {The mono-Higgs + MET signal at the Large Hadron Collider: a~study on the $\gamma \gamma $ and $b\bar{b}$ final states}.
\newblock \emph{Eur. Phys. J. C}, 82\penalty0 (10):\penalty0 914, 2022.
\newblock \doi{10.1140/epjc/s10052-022-10828-6}.

\bibitem[Belyaev et~al.(2013)Belyaev, Christensen, and Pukhov]{calchep}
Alexander Belyaev, Neil~D. Christensen, and Alexander Pukhov.
\newblock {CalcHEP 3.4 for collider physics within and beyond the Standard Model}.
\newblock \emph{Comput. Phys. Commun.}, 184:\penalty0 1729--1769, 2013.
\newblock \doi{10.1016/j.cpc.2013.01.014}.

\bibitem[Belyaev and Rojas-Abatte()]{vtdmp}
Alexander Belyaev and Felipe Rojas-Abatte.
\newblock Vector triplet dark matter with pion (vtdmp).
\newblock URL \url{https://hepmdb.soton.ac.uk/hepmdb:0820.0331}.

\bibitem[Bondarenko et~al.(2012)Bondarenko, Belyaev, Blandford, Basso, Boos, Bunichev, et~al.]{hepmdb}
M.~Bondarenko, A.~Belyaev, J.~Blandford, L.~Basso, E.~Boos, V.~Bunichev, et~al.
\newblock {High Energy Physics Model Database : Towards decoding of the underlying theory (within Les Houches 2011: Physics at TeV Colliders New Physics Working Group Report)}.
\newblock 2012.
\newblock URL \url{https://hepmdb.soton.ac.uk}.

\bibitem[Chatrchyan et~al.(2013)]{CMS:2012feb_brecosntruction}
Serguei Chatrchyan et~al.
\newblock {Identification of b-Quark Jets with the CMS Experiment}.
\newblock \emph{JINST}, 8:\penalty0 P04013, 2013.
\newblock \doi{10.1088/1748-0221/8/04/P04013}.

\bibitem[Aad et~al.(2022)]{ATLAS:2021breconstruction}
Georges Aad et~al.
\newblock {Measurement of Higgs boson decay into $b$-quarks in associated production with a top-quark pair in $pp$ collisions at $\sqrt{s}=13$ TeV with the ATLAS detector}.
\newblock \emph{JHEP}, 06:\penalty0 097, 2022.
\newblock \doi{10.1007/JHEP06(2022)097}.

\bibitem[ATL(2017)]{ATLAS:2017breconstruction}
{Variable Radius, Exclusive-k$_{T}$, and Center-of-Mass Subjet Reconstruction for Higgs($\rightarrow b\bar{b}$) Tagging in ATLAS}.
\newblock 6 2017.
\newblock URL \url{https://cds.cern.ch/record/2268678?ln=es}.

\bibitem[ATL(2015)]{ATLAS:2015breconstruction}
{Expected performance of the ATLAS $b$-tagging algorithms in Run-2}.
\newblock 7 2015.
\newblock URL \url{https://cds.cern.ch/record/2037697?ln=es}.

\bibitem[hig(2022)]{higgs_resume_2022}
{A detailed map of Higgs boson interactions by the ATLAS experiment ten years after the discovery}.
\newblock \emph{Nature}, 607\penalty0 (7917):\penalty0 52--59, 2022.
\newblock \doi{10.1038/s41586-022-04893-w}.
\newblock [Erratum: Nature 612, E24 (2022)].

\bibitem[Cepeda et~al.(2019)]{Cepeda:2019klc}
M.~Cepeda et~al.
\newblock {Report from Working Group 2}: {Higgs Physics at the HL-LHC and HE-LHC}.
\newblock \emph{CERN Yellow Rep. Monogr.}, 7:\penalty0 221--584, 2019.
\newblock \doi{10.23731/CYRM-2019-007.221}.

\bibitem[Abada et~al.(2019)]{FCC:2018vvp}
A.~Abada et~al.
\newblock {FCC-hh: The Hadron Collider}: {Future Circular Collider Conceptual Design Report Volume 3}.
\newblock \emph{Eur. Phys. J. ST}, 228\penalty0 (4):\penalty0 755--1107, 2019.
\newblock \doi{10.1140/epjst/e2019-900087-0}.

\bibitem[Pumplin et~al.(2002)Pumplin, Stump, Huston, Lai, Nadolsky, and Tung]{cteq6l}
J.~Pumplin, D.~R. Stump, J.~Huston, H.~L. Lai, Pavel~M. Nadolsky, and W.~K. Tung.
\newblock {New generation of parton distributions with uncertainties from global QCD analysis}.
\newblock \emph{JHEP}, 07:\penalty0 012, 2002.
\newblock \doi{10.1088/1126-6708/2002/07/012}.

\end{thebibliography}

\end{document}